\documentclass[10pt, a4paper]{article}

\usepackage[english]{babel}
\usepackage[utf8]{inputenc}
\usepackage{amsmath, amssymb} 
\usepackage{dsfont} % Indicator function
\usepackage[top = 1in, left = 1in, bottom = 1in, right = 1in]{geometry}
\usepackage{indentfirst} % "tab" for all new paragraphs
\usepackage[round]{natbib} % Reference settings
\usepackage{color}
\usepackage{multirow}
\usepackage{graphicx} % For tables
\usepackage{multirow} % For tables
\usepackage{diagbox} % For tables
\usepackage{makecell} % For tables
\usepackage{arydshln} % For tables
\usepackage{xcolor}
\usepackage{subcaption}
\usepackage{url}
\usepackage{enumitem}
\usepackage{caption}
\usepackage[hidelinks]{hyperref} % enables links usage
\usepackage{setspace}
\usepackage{xurl}
\usepackage{etoolbox}
\usepackage{xr}

 \usepackage[mathlines]{lineno}
\setlist{nolistsep} % remove space before starting a list

\DeclareSymbolFont{rsfs}{U}{rsfs}{m}{n}
\DeclareSymbolFontAlphabet{\mathscrsfs}{rsfs}

\newcommand{\revision}[1]{{\leavevmode\color{black}#1}}
\newcommand{\refmain}[1]{{\leavevmode\color{black}#1}}
 
\begin{document}
	% \pagenumbering{gobble}
    \pagenumbering{arabic}
	
	\begin{center}
		{\Large\text Navigating Challenges in Spatio-temporal Modelling of Antarctic Krill Abundance: Addressing Zero-inflated Data and Misaligned Covariates} \\
		\vspace{30pt}
		{\large André Victor Ribeiro Amaral${}^{1, 2, *}$, Adam M. Sykulski${}^{2}$, Sophie Fielding${}^{3}$, Emma Cavan${}^{2}$}\\
		\vspace{24pt}
		${}^{1}$University of Southampton. Southampton, UK.\\
		${}^{2}$Imperial College London. London, UK.\\
		${}^{3}$British Antarctic Survey (BAS). Cambridge, UK. \\
		${}^{*}$Corresponding author. E-mail: \texttt{\href{a.v.ribeiro-amaral@soton.ac.uk}{a.v.ribeiro-amaral@soton.ac.uk}}
		\vspace{18pt}
	\end{center}

	\begin{center}
		\textbf{Abstract}
	\end{center} \vspace{-6pt}
 
	\noindent Antarctic krill (\textit{Euphausia superba}) are among the most abundant species on our planet and serve as a vital food source for many marine predators in the Southern Ocean. In this paper, we utilise statistical spatio-temporal methods to combine data from various sources and resolutions, aiming to model krill abundance. Our focus lies in fitting the model to a dataset comprising acoustic measurements of krill biomass. To achieve this, we integrate climate covariates obtained from satellite imagery and from drifting surface buoys (also known as drifters). Additionally, we use sparsely collected krill biomass data obtained from net fishing efforts (KRILLBASE) for validation. However, integrating these multiple heterogeneous data sources presents significant modelling challenges, including spatio-temporal misalignment and inflated zeros in the observed data. To address these challenges, we fit a Hurdle-Gamma model to jointly describe the occurrence of zeros and the krill biomass for the non-zero observations, while also accounting for misaligned and heterogeneous data sources, including drifters. Therefore, our work presents a comprehensive framework for analysing and predicting krill abundance in the Southern Ocean, leveraging information from various sources and formats. This is crucial due to the impact of krill fishing, as understanding their distribution is essential for informed management decisions and fishing regulations aimed at protecting the species.
    
    \vspace{12pt}
 
	
	\newpage
	% \pagenumbering{arabic}
    \setcounter{footnote}{0} 
	
%%%%%%%%%%%%%%%%%%%%%%%%%%%%%%%%%%%%%%%%%%%%%%%%%%
\section{Introduction} \label{sec:introduction}
%%%%%%%%%%%%%%%%%%%%%%%%%%%%%%%%%%%%%%%%%%%%%%%%%%	

% \noindent\textcolor{blue}{$1^{\text{st}}$ PARAGRAPH: Frame this as a more general problem in environmental statistics.}

In environmental statistics, modelling complex ecological systems often involves substantial methodological challenges, many of which are widely encountered across various applications. A common challenge is misaligned data, where variables collected at different spatio-temporal resolutions must be integrated into a unified model. For instance, remotely sensed data, such as satellite imagery, commonly provide information on environmental phenomena at various gridded resolutions---which is fundamentally different from, e.g., data collected along transects or continuous trajectory data. Additionally, ecological datasets are often zero-inflated, containing an excess of zero observations due to the natural absence of a species or resource in certain areas. These challenges highlight the need for more sophisticated modelling frameworks that can handle such complexities, producing accurate and interpretable results while remaining computationally feasible for inference. Such aspects are central to our approach to modelling the abundance of krill in the Southern Ocean.

Antarctic krill (\textit{Euphausia superba}), hereafter referred to as ``krill,'' are one of the largest species of crustacean that lives in the water column \citep{cavan:2019} and have one of the highest biomasses of any species on Earth \citep{atkinson:2009, baron:2018, yang:2022}. Growing up to 6 cm in size and occupying a low level in the food chain, krill efficiently transfer energy by feeding on phytoplankton and serving as prey for numerous predators, including whales, seals, and penguins \citep{ruck:2014}. Their keystone role highlights their importance to the structure and functioning of the Southern Ocean ecosystem \citep{mccormack:2021}. In addition, krill are the target of the largest fishery in the region \citep{nicol:2012}. Over the past two to three decades, research on krill abundance has primarily aimed to protect krill and their predators from the impacts of fishing \citep{nicol:2012}. More recently, their role in biogeochemical cycling, particularly the carbon cycle \citep{cavan:2019}, has provided another compelling reason for conservation. Antarctic krill contribute significantly to carbon sequestration by producing long strings of carbon-rich faecal pellets that sink hundreds of metres per day, reaching deep ocean layers where the carbon can remain stored for over a century. For instance, using a combination of krill abundance data (KRILLBASE) \citep{atkinson:2017} and outputs from a physical ocean circulation model, \citet{cavan:2024} demonstrated that krill can sequester approximately 20 MtC (megatonnes of carbon) annually in the ocean interior. 

At the simplest level, protecting krill from overfishing through spatial conservation policies requires knowledge of their abundance and spatio-temporal distribution across the Southern Ocean. Although often classified as plankton, there is ongoing debate about whether krill should instead be considered ``nekton,'' as they are capable of swimming and forming massive swarms that can move against currents. As a result, while they inhabit all regions of the Southern Ocean, their distribution is highly patchy at any given time. Thus, to achieve dynamic conservation measures that adapt to the changing locations of krill, we must be able to understand their patterns in space and time. Currently, the best estimates of spatial krill biomass or abundance come from historic net haul data (KRILLBASE) and acoustic surveys, which are limited to discrete observations from research vessels \citep{fielding:2014, atkinson:2017}. The Southern Ocean's remoteness and harsh conditions restrict access to research vessels to just half the year when weather permits, making sampling both challenging and expensive. This highlights the critical need for a comprehensive modelling framework to enhance the spatial (and temporal) coverage of krill monitoring.

Integrating remotely sensed data and \textit{in situ} measurements may provide a robust approach to addressing many challenges in modelling krill abundance. Satellite imagery offers large-scale, high-resolution information on key environmental variables (e.g., sea surface temperature, chlorophyll concentration, salinity, etc.), while \textit{in situ} data provides precise, location-specific observations that capture dynamic oceanographic processes with finer detail. In this paper, the \textit{in situ} covariates come from drifters, which track near-surface ocean currents and provide valuable insights into localised water movements. These drifter trajectories allow the derivation of additional environmental covariates, such as surface speed and mass flux, as we shall detail. When combined with satellite imagery, these datasets further enhance our ability to model the physical and biological factors influencing krill distribution. To estimate krill abundance, we rely exclusively on the acoustic observations (see Section \ref{sssec:acoustic_data}), while the KRILLBASE dataset is employed for validation when extrapolating beyond the observed area. \revision{Our analysis proceeds in two directions: (I) a disaggregated spatio-temporal model that exploits the exact locations of the acoustic records, and (II) a spatial model fitted to spatio-temporally aggregated observations, aimed at predicting krill abundance across a wider area---each setting offers a distinct yet coherent view of krill distribution.}

The remainder of this paper is structured as follows. In Section \ref{sec:materials}, we introduce the krill abundance data and additional datasets used to construct covariates. Section \ref{sec:methods} outlines the spatio(-temporal) hurdle model applied to krill abundance in the South Georgia region and detail the mathematical framework for deriving spatial products from drifter trajectories. In Section \ref{sec:results}, we present and interpret the model estimates for krill biomass. Finally, in Section \ref{sec:discussion}, we provide an overall discussion of our modelling approach and findings, highlight key limitations, and suggest potential extensions for future work.

%%%%%%%%%%%%%%%%%%%%%%%%%%%%%%%%%%%%%%%%%%%%%%%%%%
\section{Materials} \label{sec:materials}
%%%%%%%%%%%%%%%%%%%%%%%%%%%%%%%%%%%%%%%%%%%%%%%%%%	

In this section, we present the datasets used in the analysis, including krill biomass measurements from acoustic and net haul data (KRILLBASE), along with remotely sensed data (e.g., satellite imagery) and \textit{in situ} measurements from drifters, which are used as covariates in our model.

\subsection{Study Area and Sampling Approach} \label{ssec:study_area} 

\begin{figure}[!ht]
	\centering
	\includegraphics[width = 1\textwidth]{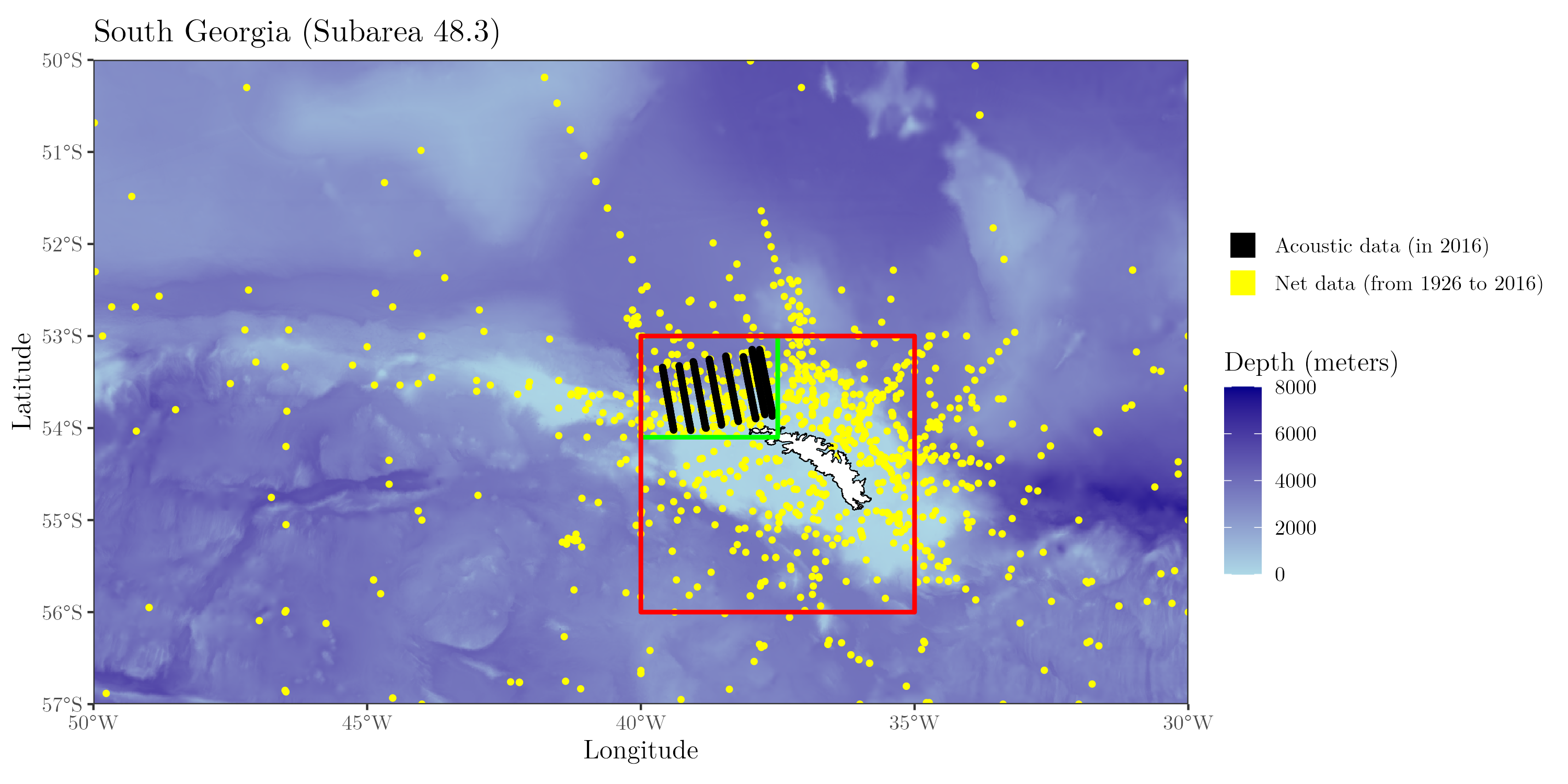}
    \caption{Study area (South Georgia, Subarea 48.3), showing sampling locations of acoustic data collected in 2016 and net haul data from KRILLBASE collected between 1926 and 2016. The green box indicates the region where acoustic data were collected, \revision{while the red box marks an area selected through visual assessment, where the net haul data were considered to offer a useful basis for comparison with model predictions} (see Section \ref{sec:results}).}
	\label{fig:main_map}
\end{figure}

Throughout this paper, we focus our analysis on subregions within the Southern Ocean, specifically around South Georgia, located in Subarea 48.3. This subarea, as defined by the CCAMLR \citep{ccamlr483:2015}, is a key ecological and management region due to its critical importance as both a krill habitat and a significant fishing area. The availability of both acoustic data and net haul data (KRILLBASE) for some parts of this region provides the necessary information to model krill abundance and distribution, making it a suitable focus for our study. Figure \ref{fig:main_map} illustrates the study area and shows the sampling locations for both acoustic (from 2016) and net haul data (spanning 1926 to 2016), and already highlights the inherent challenges of heterogeneity in the datasets, here exhibited through the highly irregular spatial sampling locations in both cases. 

Note that, although both datasets provide information on krill biomass, they do not measure it in the same manner. The net data serves only as a proxy for true krill abundance at a given location and time---since it is possible for a net to miss krill swarms even when deployed in krill-populated areas. In contrast, acoustic data enables high-resolution sampling of krill density along a vessel's path, offering a more precise measurement of krill biomass and serving as the primary data source for our analysis. In the following sections, we provide further details on these two datasets.

\subsubsection{Acoustic Data} \label{sssec:acoustic_data}

Acoustic surveys in South Georgia (Polar Ocean Ecosystem Time Series, Western Core Box) were conducted annually from 1997 to 2020, excluding the years 2002 and 2008, in intervals of 3 to 8 days within the December to February period (with December data considered as observations for the following year) \citep{fielding:2014}. Figure \ref{fig:all_pred_original_data} (left column) presents the raw data for the first and last years, while plots for the remaining years are available in \refmain{Figures SF4--SF23} (Supplementary Material). The surveys typically cover 8 transects, each 40 nautical miles in length, with a minimum separation of 10 nautical miles and a resolution of 500 metres.  

The analysis of this dataset involves several challenges. Firstly, as shown in the Figure \ref{fig:all_pred_original_data} (left column) and \refmain{Figures SF4--SF23} (Supplementary Material), many observations are zero, indicating an absence of detected krill. Secondly, krill biomass can vary substantially even over short distances; in some instances, neighbouring observations span from zero to hundreds of $\text{g}/\text{m}^2$, highlighting the spatial heterogeneity of krill biomass distribution. Lastly, while we would like to make predictions across the entire region shown in Figure \ref{fig:main_map}, our data is limited to a much smaller area (green box, Figure \ref{fig:main_map}). This limitation constrains our ability to generate reliable predictions for regions distant from the sampled locations. In Section \ref{sec:results}, we present results from two analyses covering the regions outlined by the green and red boxes (Figure \ref{fig:main_map}).

\subsubsection{KRILLBASE}

KRILLBASE is a large-scale dataset documenting the krill biomass ($\text{g}/\text{m}^2$) based on net sampling conducted throughout the Southern Ocean from 1926 to 2016 \citep{atkinson:2017}. This dataset offers valuable, long-term insights into krill abundance, which will be useful when validating our model predictions. In this paper, we pre-processed this dataset following the same procedure described in \cite{cavan:2024}, adjusting observations to estimate the expected krill density as of January each year (aligning with the acoustic data collection season), based on the collection date of each sample. Figure \ref{fig:main_map} shows the KRILLBASE sampling locations in South Georgia over the entire study period, with the corresponding spatio-temporally aggregated krill biomass, grouped into $0.2^{\circ}$ longitude by $0.125^{\circ}$ latitude cells, shown in the right-most plot of Figure \ref{fig:aggr_pred}.

\subsection{Covariates}

To effectively model krill abundance, we need to incorporate relevant covariates that capture environmental conditions influencing krill distribution. To obtain this information, we rely on multiple data sources, specifically satellite imagery and data products derived from the drifters. Satellite imagery offers large-scale, high-resolution coverage of environmental variables, while drifters provide valuable \textit{in situ} measurements of ocean currents and other local conditions.

\subsubsection{Satellite Imagery} \label{sssec:satellite_data}

We utilise ocean-related products from the Copernicus Marine Service \citep{copernicus:2024}, which provide high-resolution information on key ocean features within the study region---all of which can impact krill distribution patterns \citep{whitehouse:2009, warwick:2022}. By incorporating these satellite-derived covariates into our model, we account for large-scale environmental conditions that may drive changes in krill abundance across the South Georgia region. However, it is important to note that these datasets are not direct observations; rather, they are derived products created from satellite measurements and data processing techniques, and thus have associated uncertainty and loss of resolution due to instrumentation noise, and change of support and smoothing during processing.

Table \ref{tab:list_covariates} lists all covariates used in our analysis, including products derived from satellite imagery, drifter trajectories (see Sections \ref{sssec:drifter_data} and \ref{ssec:drifter_products}) and other environmental factors in the form of bathymetry (depth) and slope (calculated from the bathymetry). \revision{\refmain{Section SS1.1} (Supplementary Material) provides a brief description of the satellite imagery from the Copernicus Marine Service used.}

\begin{table}[!ht]
	\caption{Potential covariates for describing the spatial (and spatio-temporal) distribution of krill abundance. ${}^{\dagger}$ indicates covariates obtained from satellite imagery, and ${}^{\ddagger}$ indicates covariates derived as products from drifter trajectories. ${}^{\mathsection}$ denotes covariates observed only during the months of December, January, and February (to match the acoustic data time window). ${}^{\star}$ indicates interpolation as described in \refmain{Section SS1.2.1} (Supplementary Material).}
    \resizebox{\textwidth}{!}{%
    \centering
    \begin{tabular}{ l | l | l | l | l }
    Covariate & Spatial resolution (${}^{\circ}$) & Temporal resolution & Source & Label \\ \hline
    Bathymetry (depth)${}^{\dagger}$ & $0.01 \times 0.01$ & NA & NOAA (\href{https://doi.org/10.25921/fd45-gt74}{\texttt{10.25921/fd45-gt74}})  & \texttt{depth} \\
    Slope & $0.01 \times 0.01$ & NA & Computed based on bathymetry & \texttt{slope} \\
    Chlorophyll${}^{\dagger}$ & $0.25 \times 0.25$ & Yearly${}^{\mathsection}$ & Copernicus Marine Service (\href{https://doi.org/10.48670/moi-00019}{\texttt{10.48670/moi-00019}}) & \texttt{chlor} \\
    Potential temperature${}^{\dagger}$ & $0.083 \times 0.083$ & Yearly${}^{\mathsection}$ & Copernicus Marine Service (\href{https://doi.org/10.48670/moi-00021}{\texttt{10.48670/moi-00021}}) & \texttt{pot\_temp} \\
    Salinity${}^{\dagger}$ & $0.083 \times 0.083$ & Yearly${}^{\mathsection}$ & Copernicus Marine Service (\href{https://doi.org/10.48670/moi-00021}{\texttt{10.48670/moi-00021}}) & \texttt{salinity} \\
    Speed (satellite)${}^{\dagger}$ & $0.083 \times 0.083$ & Yearly${}^{\mathsection}$ & Copernicus Marine Service (\href{https://doi.org/10.48670/moi-00021}{\texttt{10.48670/moi-00021}}) & \texttt{speed\_sat} \\
    Surface temperature${}^{\dagger}$ & $0.05 \times 0.05$ & Yearly${}^{\mathsection}$ & Copernicus Marine Service (\href{https://doi.org/10.48670/mds-00329}{\texttt{10.48670/mds-00329}}) & \texttt{surf\_temp} \\
    Surface speed (drifters)${}^{\ddagger}$ & $0.01 \times 0.01$---after interpolation${}^{\star}$ & 1997--2020${}^{\mathsection}$ & Computed based on drifter trajectories & \texttt{speed\_drif} \\
    Expected frequency${}^{\ddagger}$ & $0.01 \times 0.01$---after interpolation${}^{\star}$ & 1997--2020 & Computed based on drifter trajectories & \texttt{expect\_freq} \\
    Residence time${}^{\ddagger}$ & $0.01 \times 0.01$---after interpolation${}^{\star}$ & 1997--2020 & Computed based on drifter trajectories & \texttt{res\_time} \\
    Mass flux${}^{\ddagger}$ & $0.01 \times 0.01$---after interpolation${}^{\star}$ & 1997--2020 & Computed based on drifter trajectories & \texttt{mass\_flux} \\
    Density of drifters${}^{\ddagger}$ & $0.25 \times 0.25$ & 1997--2020 & Computed based on drifter trajectories & \texttt{density\_drif} \\
    \end{tabular}
    }
	\label{tab:list_covariates}
\end{table}

\subsubsection{Drifter Data} \label{sssec:drifter_data}

\begin{figure}[!ht]
	\centering
	\includegraphics[width = 0.95\textwidth]{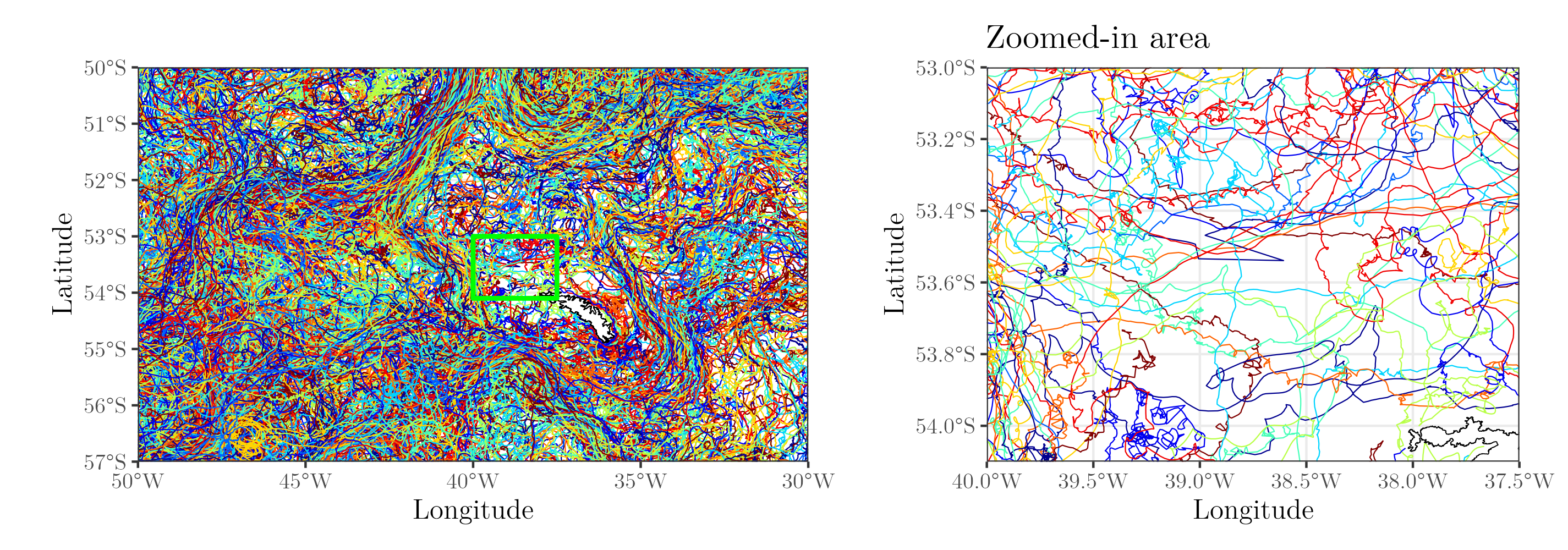}
    \caption{Left: drifter trajectories observed in South Georgia (Subarea 48.3) from 1997 to 2020, with different colours representing distinct trajectories. The green box indicates the region where acoustic data were collected, and the colours of the different trajectories are only used to ease the visualisation. Right: zoomed-in view of the green box.}
	\label{fig:drifter_trajectories}
\end{figure}

The second data source, and the most challenging to incorporate, is the drifters. Part of NOAA's (National Oceanic and Atmospheric Administration) ``Global Drifter Program,'' this dataset comprises thousands of floating buoys known as \textit{drifters} deployed in the ocean, whose positions are tracked over time by satellites, most typically using GPS. Figure \ref{fig:drifter_trajectories} (left) shows the trajectories of all drifters that were observed in the South Georgia region during the analysed time period, with a zoomed-in view of the area where the acoustic data were collected (right). These data provide valuable \textit{in situ} information about the study region which might inform krill abundance. Drifter data has previously been used to inform abundance and dynamics of a broad range of ocean-borne species and objects \citep{o2021estimating}, including plankton \citep{laso2023holistic}. While krill are not like plankton and can swim against weak currents, the impacts of ocean dynamics and currents on krill abundance and krill flux is nonetheless well documented \citep{murphy2004southern}, therefore, there is reasonable scientific rationale for drifter data being informative in predicting krill abundance.

In this paper, we consider all the trajectories presented in Figure \ref{fig:drifter_trajectories} (left). Specifically, the positions of the buoys are recorded on an hourly basis \citep{elipot2016global}, with a total of 1,294 trajectories observed from 1997 to 2020. These trajectories vary in length from 122 to 8,797 points, adding up to 1,475,178 unique observations (or approximately 168.4 years' worth of data).

However, the drifter trajectories are not yet ready-to-use covariates as they are stored in the form of timestamped trajectories (as in Figure~\ref{fig:drifter_trajectories}) rather than gridded spatio(-temporal) products as would be typical from e.g., satellite imagery. To proceed, we therefore aim to transform the drifters into spatially gridded data by extracting specific features of interest from them, which we shall describe in detail in Section \ref{ssec:drifter_products}.  As we will show, various spatial data products can be derived from drifter trajectories, providing potentially orthogonal information to satellite imagery and enhancing our model. While drifter data have previously been applied in krill abundance modelling \citep{siegel:2013}, some of the products introduced in this study represent a novel use of this dataset as environmental covariates to describe krill distribution.

%%%%%%%%%%%%%%%%%%%%%%%%%%%%%%%%%%%%%%%%%%%%%%%%%%
\section{Methods} \label{sec:methods}
%%%%%%%%%%%%%%%%%%%%%%%%%%%%%%%%%%%%%%%%%%%%%%%%%%	

In this section, we outline the modelling framework and inference approach used to model krill abundance in the South Georgia region, as well as the methods for deriving valuable products from drifter trajectories.

\subsection{Spatio-temporal Modelling} \label{ssec:spatio-temporal-model}

Throughout this paper, we use a Hurdle-Gamma model \citep{cragg:1971, min:2022} to address the challenges in modelling krill abundance, where the data consist of a non-negative continuous outcome with excess zeros. The Hurdle-Gamma model is particularly useful in this context, as it jointly models the probability of krill absence (i.e., presence-absence) and the distribution of non-zero abundance values.

Let $\mathcal{X} \subset \mathbb{R}^2$ denote the continuous spatial domain, with observed locations $(s_1, \cdots, s_n) \subset \mathcal{X}$. \revision{Similarly, we define $\mathcal{T}$ as the temporal domain, with $t \in \{1, \cdots, T\}$ indexing discrete time points.} Following the notation in \cite{krainski:2018}, let
\begin{align*}
    z_{it} =
        \begin{cases}
        1, & \text{ if the krill biomass is non-zero at  location } s_i \text{ at time } t \\
        0, & \text{ otherwise}
        \end{cases}
\end{align*}
and $y_{it}$ denotes the krill biomass at location $s_i$ at time $t$, given that the biomass is non-zero. Specifically, we model the presence-absence component as $z_{it} \sim \text{Bernoulli}(\pi_{it})$ and the positive biomass as $y_{it} \sim \text{Gamma}(a_{it}, b_{it})$. The Gamma distribution is parametrized such that, $\mathbb{E}(y_{it}) = \mu_{it} = a_{it} / b_{it}$ and $\text{Var}(y_{it}) = a_{it} / b_{it}^2$. 

The linear predictors for the presence-absence indicator $z_{it}$ and the positive biomass $y_{it}$ are specified as follows
\begin{align} \label{eq:linear-predictor-z}
    \text{logit}(\pi_{it}) = \beta_0^z + \beta_1^z \texttt{cov}_{1, it}^z + \cdots + \beta_{\ell_1}^z \texttt{cov}_{\ell_1, it}^z + \psi_{it},
\end{align}
and
\begin{align} \label{eq:linear-predictor-y}
    \log(\mu_{it}) = \beta_0^y + \beta_1^y \texttt{cov}_{1, it}^y + \cdots + \beta_{\ell_2}^y \texttt{cov}_{\ell_2, it}^y + \gamma \cdot \psi_{it} + \xi_{it},
\end{align}
where $\ell_1$ and $\ell_2$ denote the number of covariates in each model, which may overlap, and $\psi_{it}$ and $\xi_{it}$  are spatio-temporal random effects. Here, $\gamma$ serves as a ``copy'' factor to scale the shared random effect $\psi_{it}$ in the biomass model, allowing for dependencies between the presence-absence and biomass components. \revision{We note that, although the covariate effects are specified linearly in Equations \eqref{eq:linear-predictor-z} and \eqref{eq:linear-predictor-y}, they could be replaced by mildly non-linear bases---such as cubic B-splines with a small knot set \citep{fahrmeir:2004}---if future analyses indicate a clear benefit.}

For the random effects, we define $\psi_{it}$ (similarly, $\xi_{it}$) as an autoregressive process to capture temporal correlation while allowing for spatial dependency at each time point \citep{moraga:2019}. In particular, we set
\begin{align} \label{eq:AR1-GP}
    \psi_{it} = \alpha_{\psi} \psi_{i, (t - 1)} + \phi_{it},
\end{align}
where $|\alpha_{\psi}| <1$, $\psi_{i1} \sim \text{Normal}(0, \sigma^2_{\phi}/(1 - \alpha_{\psi}^2))$, and $\phi_{it}$ is a temporally independent but spatially dependent Gaussian Process (GP) at each year with covariance given by a Matérn kernel, i.e.,
\begin{align} \label{eq:matern-model}
	\text{Cov}(\phi_{it}, \phi_{jt}) = \frac{\sigma^2_{\phi}}{2^{\nu - 1}\Gamma(\nu)} (\kappa \cdot h)^{\nu} K_{\nu}(\kappa \cdot h),
\end{align}
where $h = ||s_{it} - s_{jt}||$ is the Euclidean distance between the locations $s_{it}$ and $s_{jt}$, and $\sigma^2_{\phi}$ denotes the marginal variance. $\Gamma(\cdot)$ is the Gamma function, and $K_{\nu}(\cdot)$ is a modified Bessel function of the second kind, such that $\nu >0$ determines the mean square differentiability of the corresponding process. Lastly, $\kappa > 0$ is related to the range $\rho$, such that $\rho = \sqrt{8\nu}/\kappa$.

Finally, while we assume Gaussianity for the spatial field, this may not fully capture high spatial heterogeneity in the data, potentially leading to oversmoothing of distinct features such as sharp valleys or peaks. To address this, we perform a sensitivity analysis to assess the robustness of our results under this assumption before drawing any conclusions (see discussion in Section \ref{sec:results}). This approach balances interpretability and computational feasibility, ensuring that the model remains practical to fit without imposing an excessive computational burden (Section \ref{sssec:inference}).

\subsubsection{Spatial Modelling} \label{sssec:spatio-model}

In Section \ref{ssec:aggregated}, we perform an aggregated spatial analysis and, in this instance, drop the temporal component from Equations \eqref{eq:linear-predictor-z} and \eqref{eq:linear-predictor-y}. In this setting, the corresponding spatial Hurdle-Gamma model will be defined as before; however, the random effects $\psi_i$ and $\nu_i$ will be modelled as Gaussian processes in space only, using a Matérn kernel similar to that in Equation \eqref{eq:matern-model}. 

\revision{This spatial-only formulation can be interpreted as an empirical approximation to the mean (averaged over time) of the full spatio-temporal model; i.e., after averaging the original response $z_{it} \cdot y_{it}$ over the survey years, its model-based expectation $(1/T)\sum_{t}\pi_{it}\cdot\mu_{it}$ can be approximated by the product of the time-averaged components $\bar{\pi}_{i} \cdot \bar{\mu}_i$. That simplification relies on three working conditions---(I) covariate effects do not vary with year, (II) no strong residual year trend remains once covariates are included, and (III) the underlying biomass field shows no long-term drift—under which temporally averaged covariates and a purely spatial random effect provide a coherent large-scale picture of krill distribution. However, if any of these conditions is violated, a model that retains an explicit time dimension would be preferable. In \refmain{Section SS2} (Supplementary Material), we illustrate this equivalence by means of a simulation study.}

\subsubsection{Inference} \label{sssec:inference}

Inference is conducted within a Bayesian framework using Integrated Nested Laplace Approximations (INLA) \citep{rue:2009} to efficiently approximate posterior distributions in latent Gaussian models, which is particularly advantageous for complex spatio-temporal structures (as in Section \ref{ssec:spatio-temporal-model}). Model fitting also relies on the Stochastic Partial Differential Equation (SPDE) approach, where the Gaussian field with a Matérn covariance structure is expressed as the solution of a SPDE \citep{whittle:1963} and then approximated by a Gaussian Markov Random Field (GMRF) on a triangulated mesh \citep{lindgren:2011}, enabling a scalable representation of spatial dependence. Finally, we use Penalised Complexity (PC) priors \citep{simpson:2017} for the parameters in the random effects, following the recommendations of \cite{krainski:2018}. In practice, we implement our models using \texttt{R-INLA} \citep{lindgren:2015}, and the corresponding code is available at \url{https://github.com/avramaral/krill_abundance}.

\subsection{Deriving Products from Drifter Trajectories} \label{ssec:drifter_products}

In this section, we use the drifter trajectory data introduced in Section \ref{sssec:drifter_data} to derive spatial products for use as covariates in our krill abundance model. We note in passing that these products may also be valuable for predicting other ocean phenomena, such as the spread of oil spills, plankton, and plastic pollution.

We begin by establishing some notation. The observed position of drifter $i$ in a spatial region of interest $\mathcal{X}$ at time $t$ will be denoted by $q_i(t) \in \mathcal{X}$, representing its latitude-longitude coordinates. The collection of consecutive positions for each drifter $i$ observed in region $\mathcal{X}$ will be denoted by $\{q_i(t)\}$ and is known as the \textit{trajectory} of drifter $i$. Note that if the drifter leaves the spatial region of interest $\mathcal{X}$, but then re-enters, then multiple trajectories may be collected from the same drifter, and for simplicity we will denote each such trajectory with its own drifter index value $i$.

A primary use of drifter trajectory data is to track the velocity of the drifter along its path---often referred to in fluid dynamics as the \textit{Lagrangian} velocity, named so because the drifter is deliberately designed to mimic a buoyant particle as it moves through time and space and thus has a Lagrangian perspective of the horizontal fluid flow near the surface. There are many works focussed on deriving statistics from Lagrangian velocities, see e.g., \cite{lacasce2008statistics,sykulski2016lagrangian}, where we employ similar notation and modelling principles here.
As is typical in ocean flow analysis, the Lagrangian velocity of drifter $i$ at time $t$ will be modelled in the complex plane by $z_i(t) = u_i(t) + \mathrm{i}v_i(t)$ where $u_i(t)$ and $v_i(t)$ correspond to the zonal (eastward) and meridional (northward) velocities respectively, and are obtained in practice from $\{{q}_i(t)\}$ by some form of differencing or gradient modelling over time for each drifter $i$ \citep{elipot2016global}. Representing two-dimensional time series in the complex plane is common in signal processing applications, especially when the two dimensions are measuring the same quantity (in this case, velocities) in orthogonal directions, and offers computational and modelling advantages over vector or bivariate representations, as reviewed in \cite{sykulski2017frequency}, and as we shall take advantage of here.

Therefore we have at our disposal a collection of trajectories $\{{q}_i(t)\}$ and corresponding velocities $\{z_i(t)\}$ for drifter $i$ inside region $\mathcal{X}$, where in Section~\ref{sec:results} the region $\mathcal{X}$ will be the entire Subarea 48.3 shown earlier in Figure~\ref{fig:main_map}. We now seek to derive or ``engineer'' spatial covariates from $\{{q}_i(t),z_i(t)\}$ that can be utilised in our Hurdle-Gamma model of Equations~\eqref{eq:linear-predictor-z} and \eqref{eq:linear-predictor-y}. The key opportunity in deriving such covariates is to capture the local information content inherent in drifter trajectories (and their velocity gradients) that cannot be captured from satellite imagery. We now propose five such covariates, as shown in Table~\ref{tab:list_covariates}, which each capture different characteristics of the drifter data.

\begin{figure}[!ht]
	\centering
	\includegraphics[width = 1\textwidth]{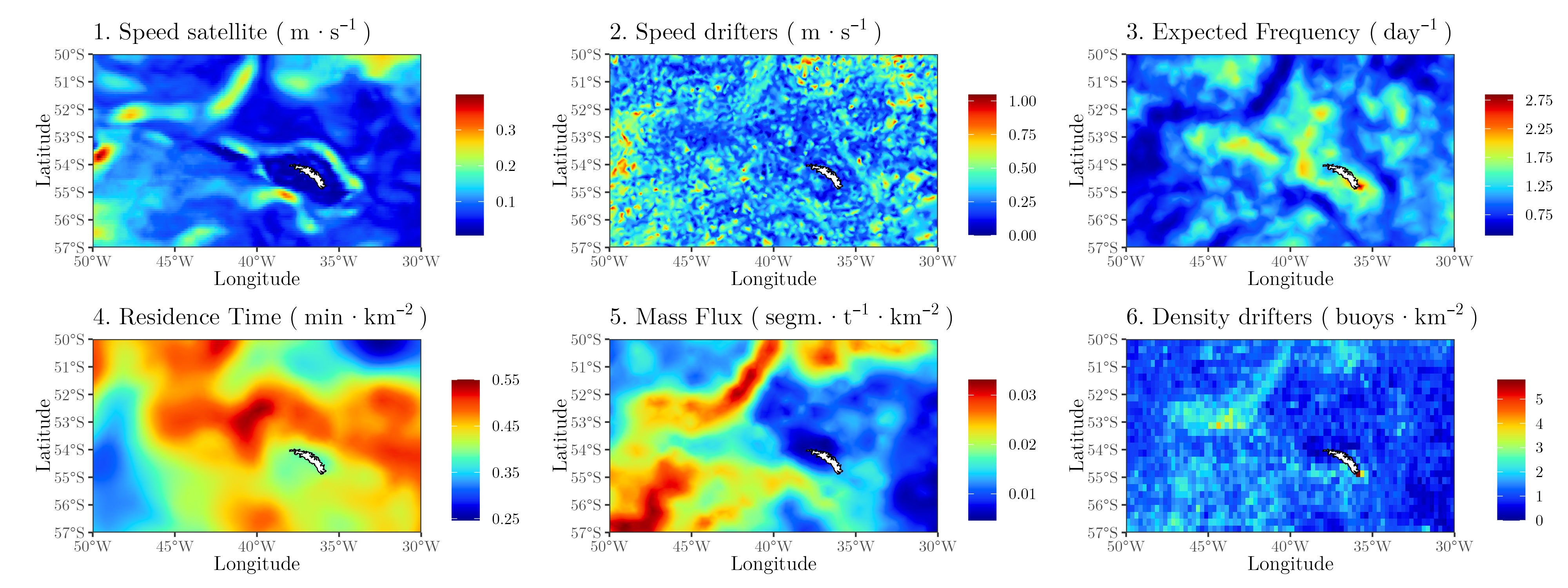}
    \caption{Covariates in South Georgia (Subarea 48.3), as described in Table \ref{tab:list_covariates}. All drifter-derived products were computed based on the trajectories collected from 1997 to 2020. 1: Speed ($\text{m} \cdot \text{s}^{-1}$) from the satellite, averaged over 1997--2020. 2: Surface speed ($\text{m} \cdot \text{s}^{-1}$) from the drifters. 3: Expected frequency ($\text{day}^{-1}$). 4: Residence time ($\text{min} \cdot \text{km}^{-2}$). 5: Mass flux ($\text{segments} \cdot \text{t}^{-1} \cdot \text{km}^{-2}$), where ``time'' refers to the entire observational period, i.e., 24 years. 6: Density of drifters ($\text{buoys} \cdot \text{km}^{-2}$).
}
	\label{fig:drifter_products}
\end{figure}

\begin{itemize}
\item \textbf{Surface speed (drifters):} we compute the speed of all drifter observations given by $|z_i(t)|$ and then map these to their corresponding locations ${q}_i(t)$. After which we create a spatially gridded product at the desired resolution by interpolating using Gaussian processes with a Mat\'ern kernel, as detailed in \refmain{Section SS1.2.1} (Supplementary Material). Note, importantly, that this covariate is expected to be different from the speed (satellite) covariate in Table~\ref{tab:list_covariates}, as the satellite data we use provides estimates of the \textit{geostrophic} velocity computed from sea surface height (SSH) gradients, and is averaged over different depths, whereas drifter speeds are expected to be a mix of geostrophic and ageostrophic velocities (caused for example by surface winds) at or near the surface. 
This difference is explained in detail by \cite{o2023probabilistic}. 
% Hence, drifter speeds are expected to be markedly different from satellite-derived geostrophic speeds, as is well documented, see e.g., \cite{o2023probabilistic}.
Figure \ref{fig:drifter_products} shows maps of the surface speed estimates in Subarea 48.3 from satellite data and drifter observations for comparison.

\item \textbf{Expected frequency:} the speed from drifters is potentially informative, but ignores the information contained in the \textit{shape} of the drifter trajectories $\{{q}_i(t)\}$. As detailed in Section 2.7 of \cite{lacasce2008statistics}, and Section 2.2 of \cite{lilly2017fractional}, one of the best ways to understand the shape of drifter trajectories is via the  \textit{Lagrangian frequency spectrum} defined by
\begin{align*} 
    S_z(\omega) = \int_{-\infty}^\infty s_z(\tau) e^{-\mathrm{i}\omega\tau} d\tau, \quad \omega\in\mathbb{R},
\end{align*}
where $s_z(\tau)$ is the autocovariance of the complex-valued velocity process $z(t)$ given by
\begin{align*}
    s_z(\tau) = \mathbb{E}(z(t)z^\ast(t+\tau))-\mathbb{E}(z(t))\mathbb{E}(z^\ast(t)), \quad \tau\in\mathbb{R},
\end{align*}
where $z(t)$ is a second-order stationary stochastic process such that $s_z(\tau)$ is invariant over time $t$, and $z^\ast(t)$ denotes the complex conjugate of $z(t)$. The Lagrangian frequency spectrum can therefore be interpreted as the power spectral density of the velocity process, as it decomposes the second-order variability, or \textit{power}, of the velocity process by frequency. Drifters that have a tendency to oscillate or jitter will have more power at high frequencies, and drifters that have a tendency to move in straighter lines will have more power at low frequencies. An informative covariate that summarises this content is the \textit{expected frequency} of the velocity process given by
\begin{align} \label{EF}
    \text{EF}_z = \int_{-\infty}^\infty \frac{|\omega|S_z(\omega)}{\int_{-\infty}^\infty S_z(\omega) d\omega} d\omega,
\end{align}
where the density $S_z(\omega)$ above has been normalised to integrate to 1 (such that it can be interpreted as a probability density over $\omega$ in some sense) thus explaining the term ``expected frequency.'' 
%To arrive at the final expression in Equation \eqref{EF}, we have used the inverse Fourier transform of Equation \eqref{FT}, namely
% \begin{align*}
%     s_z(\tau) = \frac{1}{2\pi}\int_{-\infty}^\infty S_z(\omega) e^{\mathrm{i}\omega\tau} d\omega, \quad \omega\in\mathbb{R},
% \end{align*}
% where $s_z(0) = \sigma^2$ is the variance of the process $z(t)$. 
In practice, we have at our disposal sampled velocity time series $z_i(t)=\{z_i(t_1),\ldots,z_i(t_{n_i})\}$ for each drifter $i$ (of length $n_i$). Here, we can approximate the Lagrangian frequency spectrum via a tapered spectral estimate as follows
\begin{align} \label{periodogram}
    \hat{S}_i(\omega) = \frac{\Delta}{n_i}\left|\sum_{j=1}^{n_i} h_jz_i(t_j) e^{-\mathrm{i}j\omega\Delta}\right|^2,
\end{align}
where $\Delta$ is the temporal sampling interval and is assumed constant (which it is with the drifter data used in this paper; see Section \ref{sssec:drifter_data}). The sequence $\{h_j\}$ in Equation \eqref{periodogram} is known as a data taper that satisfies $\sum_{j=1}^{n_i} h_j^2 =1$, where we select $\{h_j\}$ to be a DPSS (discrete prolate spheroidal sequence) of order 1 (with bandwidth parameter set to 4), and is used to remove bias in the estimate of the spectrum, see \citet[Chapter 6]{percival1993spectral} for more details. We can then approximate the expected frequency in Equation \eqref{EF} by
% \begin{align*}
%     \widehat{\text{EF}}_i = \frac{1}{n_i\Delta\hat\sigma^2}\sum_{k=1}^{n_i}|\omega_k|\hat{S}_i(\omega_k),
% \end{align*}
\begin{align*}
    \widehat{\text{EF}}_i = \frac{1}{\kappa}\sum_{k=1}^{n_i}|\omega_k|\hat{S}_i(\omega_k),
\end{align*}
where
\begin{align*}
    (\omega_1,\ldots,\omega_{n_i}) = \frac{2\pi}{n_i\Delta}(-\lceil{n_i/2}\rceil+1,\ldots,-1,0,1,\ldots,\lfloor{n_i/2}\rfloor),
\end{align*}
are the observed Fourier frequencies and
%$\hat\sigma^2$ is the sample variance of $z_i(t)$, 
$\kappa = \sum_{k = 1}^{n_i} \hat{S}_i(\omega_k)$. %DPSS was obtained using the \texttt{dpss.taper()} function from the \texttt{waveslim} package \citep{waveslim:2024}, with parameters $\texttt{k} = 1$ (number of tapers) and $\texttt{nw} = 4$ (time-bandwidth parameter).
The calculation of expected frequency requires the velocity time series to be approximately stationary, which will not generally be the case for an entire drifter trajectory in our region of interest. Therefore, we compute the expected frequency for each drifter trajectory over temporal windows (with 50\% overlap) of length 5 days, which is considered to be a good approximation of the ``decorrelation timescale" (i.e., the timescale at which a drifter ``forgets" its history of movement), and is a standard choice in ocean drifter analysis (see \cite{o2021estimating} and references therein). Finally, we derive a spatial gridded map by taking the set of computed expected frequencies and mapping them onto the midpoint location of each trajectory segment and then spatially smoothing onto a grid using Gaussian processes, as detailed in \refmain{Section SS1.2.1} (Supplementary Material).

\item \textbf{Residence time:} the expected frequency summarises the non-zero frequency content of a drifter trajectory. On the other hand, the zero frequency of the Lagrangian frequency spectrum yields a quantity known as the \textit{diffusivity} which from Section 2.3 of \cite{lilly2017fractional} can be related via several quantities such that
\begin{equation}
    \kappa_z = \frac{1}{4}S_z(0) = \frac{1}{4}\int_{-\infty}^\infty s_z(\tau) d\tau = \lim_{t \rightarrow \infty} \frac{1}{4} \frac{d}{dt} \mathbb{E}\left\{|q(t)|^2\right\},
    \label{diffusivity}
\end{equation}
where $q(t)=\int_0^t z(\tau)d\tau$ is the (complex-valued) displacement of the drifter at time $t$ where $z(t)$ is a zero-mean velocity process. The diffusivity can therefore also be understood as the integral of the autocovariance sequence over all lags, or as the expected rate of change over time of the squared displacement of the drifter after its mean is removed (i.e., the rate of diffusion)---thus linking Equation \eqref{diffusivity} to the physical notion and definition of diffusivity. Therefore, we propose a covariate from the drifters which can capture spectral information missing in the expected frequency, namely the diffusivity. However, diffusivity is difficult to estimate individually from single drifter trajectories, as the spatio-temporally varying local mean velocity (also known as the mean flow) must be removed and separated \citep{oscroft2020separating}, but in the Global Drifter Program the local mean flow is in general unknown due to drifter sparsity. We therefore instead estimate a quantity known as the {\em residence time}, commonly used in fluid dynamics and chemistry \citep{nauman2008residence}, which estimates how long a fluid particle spends within a control volume of fixed size, thus incorporating both diffusivity and mean flow features. Specifically, in our case, the residence time is estimated by dividing the spatial region into overlapping circular windows of constant radius. Then, within each circle, we compute the median length of time a drifter trajectory consecutively remains inside the circle as our estimate of the residence time. We then map onto a spatial grid as with surface speed (drifters) and expected frequency. Further implementation details specific for the krill analysis and Subarea 48.3 can be found in \refmain{Section SS1.2} (Supplementary Material).

\item \textbf{Mass flux:} the residence time computes the \textit{average time} a drifter continuously spends in a fixed spatial region. A natural orthogonal covariate to also include is the \textit{number} of drifters that pass through this region over time. This can be interpreted as the \textit{mass flux} of drifters as it measures the rate at which drifters move across a unit area per unit of time. The motivation to include this covariate also comes from \cite{murphy2004southern}, who find associations between water volume flux (which the drifters are mimicking near the surface) and krill flux. For our analysis, mass flux is computed in exactly the same way as residence time, see \refmain{Section SS1.2} (Supplementary Material) for details.

\item \textbf{Density of drifters:} lastly, as the drifters are freely floating then they are not uniformly sampling the ocean and instead are likely to be preferentially sampling the ocean due to the impact of, for example, convergent or divergent zones \citep{middleton1986kinematic}. Although the density of drifters and krill will not necessarily aggregate in the same way, it nonetheless could be informative as a covariate. We therefore include a basic estimate of the density of drifters which corresponds to the total number of hours spent by drifters in each pixel of the spatial image, as detailed in \refmain{Section SS1.2} (Supplementary Material).

\end{itemize}

The five proposed drifter products are plotted in Figure \ref{fig:drifter_products} for Subarea 48.3. Additionally, we include a spatial plot of the ``speed (satellite)'' covariate for comparison, which, as expected, shows related drifter speeds but also reveals some differing structures. While the five drifter products are clearly not entirely orthogonal (e.g., the mass flux is higher in regions of increased speed, as expected), none of them appear to be collinear. Thus, considering the rich information content of the drifter data, with approximately 1.5 million unique observations, we incorporate all these products into our spatio(-temporal) analysis of krill abundance in the next section. However, we emphasise that in other applications, it may be more appropriate to include only a subset of these products. For instance, in Section \ref{ssec:aggregated}, we applied stepwise forward selection and added only residence time and mass flux from the drifters, in addition to other satellite-based covariates. Lastly, it is worth noting that drifters are attached to a drogue (also known as a sea anchor) and measure near-surface currents at approximately 15 meters below the water's surface, whereas krill swarms can occur at greater depths. Consequently, while the relationship between some of the maps in Figure \ref{fig:drifter_products} and krill abundance may be significant, it might not be as strong across all locations.

%%%%%%%%%%%%%%%%%%%%%%%%%%%%%%%%%%%%%%%%%%%%%%%%%%
\section{Results} \label{sec:results}
%%%%%%%%%%%%%%%%%%%%%%%%%%%%%%%%%%%%%%%%%%%%%%%%%%	

Following the modelling framework described in Section \ref{ssec:spatio-temporal-model} and incorporating covariates from satellite imagery and drifter products (Table \ref{tab:list_covariates}), we fit a Hurdle-Gamma model for krill biomass from acoustic data \revision{only} (Section \ref{sssec:acoustic_data}) under two settings. First, we apply the model to the disaggregated data at its original spatio-temporal resolutions and focus on the region where we observed the data (green box, Figure \ref{fig:main_map}). In this setting, our primary interest lies in the interpretability of some model parameters. Second, to enhance predictive capability outside the observed window (red box, Figure \ref{fig:main_map}) and in line with the approach of \cite{warwick:2022}, we fit a spatial-only version of our model to the acoustic data aggregated across space and time. This setup also enables variable selection at a feasible computational cost. 

\subsection{Disaggregated Spatio-temporal Modelling} \label{ssec:disaggregated}

First, we fit the spatio-temporal Hurdle-Gamma model introduced in Section \ref{ssec:spatio-temporal-model}, \revision{retaining the complete random-effect structure and} including all covariates listed in Table \ref{tab:list_covariates} in the linear predictors for both the presence-absence and positive biomass components---i.e., Equations \eqref{eq:linear-predictor-z} and \eqref{eq:linear-predictor-y}, respectively. Full details on this model are given in \refmain{Section SS3.1} (Supplementary Material). This approach was chosen to avoid the need for multiple model re-fits, as the associated computational cost was prohibitively high, despite the optimised inference specifications detailed in Section \ref{sssec:inference}. \revision{Lastly, although the formulation in Section \ref{ssec:spatio-temporal-model} is well-defined, note that practical identifiability issues between the two spatio-temporal random effects can arise when data are insufficiently informative or when priors are poorly specified.}

\revision{In this setting, we focus on characterising the presence-absence component, whose linear predictor is defined in Equation \eqref{eq:linear-predictor-z}. Furthermore, as previously mentioned, we examine key hyperparameters of the spatio-temporal random effects to gain deeper insight into the structure and design of the acoustic surveys' sampling strategy.}

\begin{figure}[!ht]
	\centering
	\includegraphics[width = 1\textwidth]{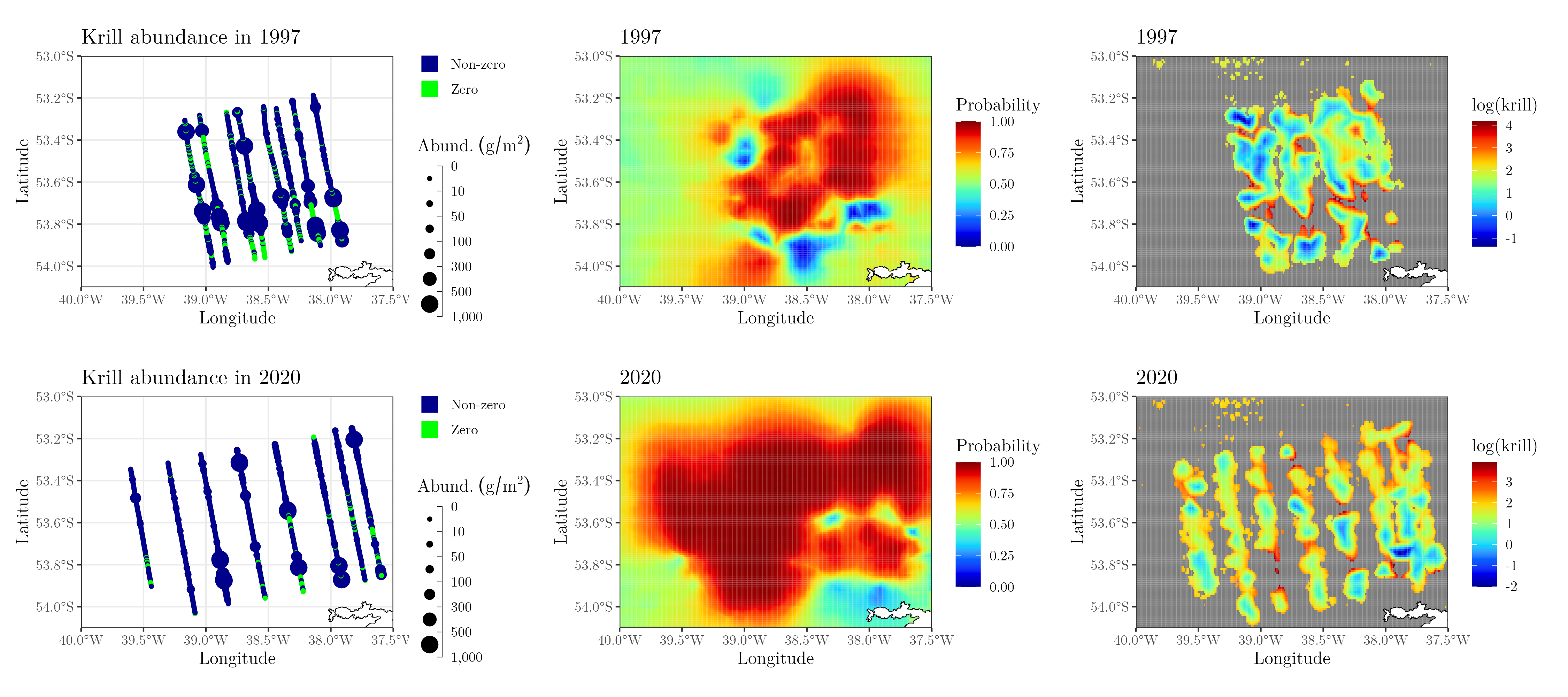}
    \caption{Left column: observed acoustic krill biomass data. Middle column: estimated probability of non-zeros. Right column: predicted krill biomass ($\text{g}/\text{m}^2$), with predictions having a standard deviation greater than 3 (on the log scale) being masked out. \revision{The two rightmost columns are based on the mean of the corresponding predictive distributions.}}
	\label{fig:all_pred_original_data}
\end{figure}

\refmain{Table ST3} (Supplementary Material) shows the estimated coefficients, and Figure \ref{fig:all_pred_original_data} shows the predicted values for both presence-absence and positive biomass components in 1997 and 2020. The corresponding results for the remaining years are shown in \refmain{Figures SF4--SF23} (Supplementary Material). In the right-most plots of Figure \ref{fig:all_pred_original_data}, we masked out predicted values at locations with high uncertainty---specifically, where the standard deviation is greater than 3 (on the log scale, or approximately $20 \text{g}/\text{m}^2$). As previously noted, the variability in the disaggregated data makes any spatio-temporal extrapolation beyond the observed locations extremely challenging using the modelling framework from Section \ref{ssec:spatio-temporal-model}. This also explains our focus on the region delineated by the green box (Figure \ref{fig:main_map}). To address such a limitation and improve our ability to make predictions in non-observed areas, we shift to an aggregated analysis in Section \ref{ssec:aggregated}.

\begin{figure}[!ht]
	\centering
	\includegraphics[width = 1\textwidth]{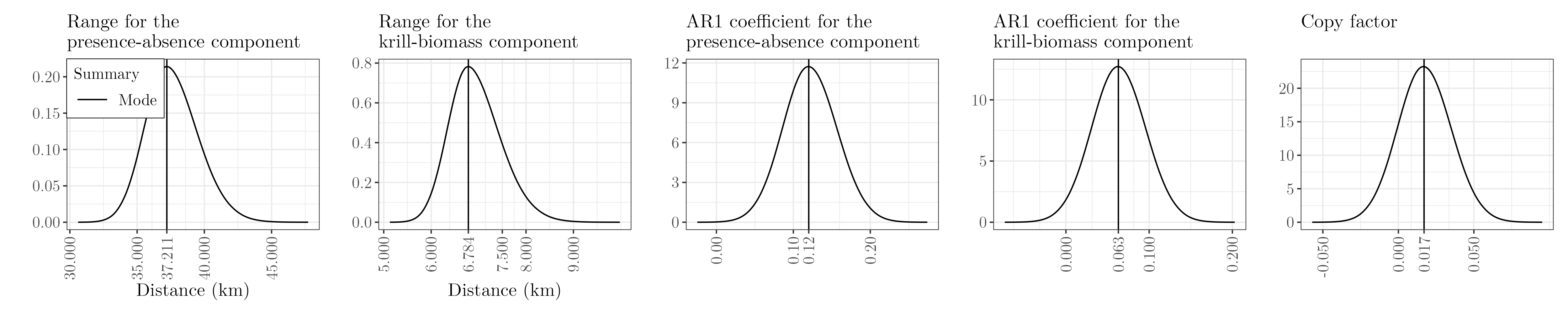}
    \caption{Posterior distributions for the range, AR1 coefficients from Equation \eqref{eq:AR1-GP}, and ``copy'' factor $\gamma$ from Equation \eqref{eq:linear-predictor-y}.}
	\label{fig:all_post}
\end{figure}

In addition to the predicted processes shown in Figure \ref{fig:all_pred_original_data}, we may also be interested in interpreting certain model hyperparameters, particularly those related to the estimated random effects. Figure \ref{fig:all_post} presents the posterior distributions of key parameters, including the range $\rho$, as in Equation \eqref{eq:matern-model}, for the random effects both in the presence-absence linear predictor and in the positive krill biomass linear predictor, i.e., Equations \eqref{eq:linear-predictor-z} and \eqref{eq:linear-predictor-y}, respectively.

Using the mode of the corresponding posterior distribution as a point estimate for the range, we find that, in the presence-absence component, it is approximately 20 nautical miles (approx. $37$ km), whereas in the positive krill biomass component, it is approximately 3.7 nautical miles (approx. $6.8$ km). As noted in Section \ref{sssec:acoustic_data}, the transects are positioned at least 10 nautical miles apart to ensure independent samples across different transects. In this context, our estimates could further refine sampling routes for future surveys, as $\rho = \sqrt{8\nu}/\kappa$ indicates the distance at which spatial correlation is close to $0.1$ \citep{cameletti:2013}. 

However, before using these estimates to guide adjustments in data collection strategies, it is essential to assess their robustness under the Gaussianity assumption for the latent field. In \refmain{Section SS3.1.1} (Supplementary Material), we conducted a sensitivity analysis by re-estimating the hyperparameters for observations generated from a latent non-Gaussian model. The results suggest that, while a potentially misspecified model may introduce a small bias in the range parameter (in particular, in our experiment, we noticed an upward bias of 5-10\% for the parameter $\kappa$), the overall conclusions regarding acoustic survey sampling remain unaffected, as these differences are not substantial enough to meaningfully impact interpretation. 

\subsection{Aggregated Spatial Modelling} \label{ssec:aggregated}

\begin{figure}[!ht]
	\centering
	\includegraphics[width = 1\textwidth]{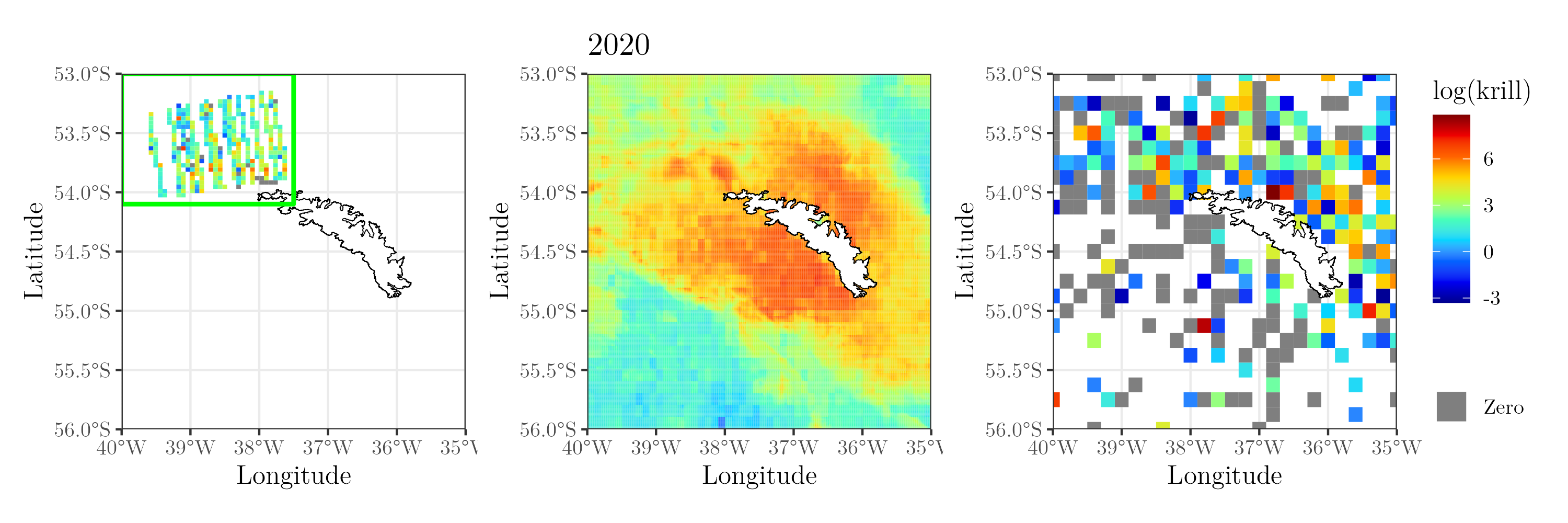}
    \caption{Left column: aggregated acoustic krill biomass data. Middle column: predicted krill biomass in 2020 \revision{(based on the mean of the predictive distribution)}. Right column: KRILLBASE (net haul data), aggregated temporally and spatially. Krill biomass is in $\text{g}/\text{m}^2$.}
	\label{fig:aggr_pred}
\end{figure}

Following the approach of \cite{warwick:2022}, who estimated krill abundance in the northern Antarctic Peninsula region, we fit a spatial Hurdle-Gamma model (as stated in Section \ref{sssec:spatio-model}) to an aggregated version of the original acoustic krill data. Specifically, we aggregate the acoustic data both temporally \revision{(all data from 1997 to 2020, Section \ref{sssec:acoustic_data})} and spatially \revision{by taking the mean of observations within cells} measuring $0.067^{\circ}$ longitude by $0.036^{\circ}$ latitude, corresponding to approximately $4 \times 4$ km (as shown in the left column of Figure \ref{fig:aggr_pred}). Modelling this aggregated version of the data reduces issues of high spatial variability over short distances, making the Gaussianity assumption more reasonable and decreasing uncertainty in predictions beyond the observed area. While this approach sacrifices spatio-temporal resolution, \revision{it is explicitly aimed at providing a broad, large-scale picture of krill biomass across a wider area.} Thus, in this section, we focus on making predictions within the red box (Figure \ref{fig:main_map}), where there are more observations from KRILLBASE (net haul data), enabling us to compare and evaluate the accuracy of our extrapolated predictions---\revision{although, as discussed in Section \ref{ssec:study_area}, the KRILLBASE covers a different temporal range compared to the acoustic krill data (see \refmain{Figure SF26}, Supplementary Material). Additionally, the net haul data do not measure krill biomass in the same way as the acoustic surveys and thus serve only as a proxy for the true spatial distribution.}

In this scenario, since the model is computationally much cheaper to fit, we can perform variable selection. Specifically, we perform stepwise forward selection based on the Watanabe-Akaike Information Criterion (WAIC) \citep{watanabe:2013, gelman:2014}. Additionally, we tested alternative models with simpler random effect structures (also using stepwise forward variable selection). However, the original model, i.e., the spatial Hurdle-Gamma model with linear predictors as in Equations \eqref{eq:linear-predictor-z} and \eqref{eq:linear-predictor-y}, consistently outperformed these alternatives (\refmain{Section SS3.2.1}, Supplementary Material), reinforcing our choice to use it. Full details on the selected model are provided in \refmain{Section SS3.2}, where \refmain{Table ST5} (Supplementary Material) shows the estimated coefficients. The covariates included in the linear predictors, in addition to the intercept, were as follows: for the presence-absence component, as in Equation \eqref{eq:linear-predictor-z}, we considered chlorophyll, potential temperature, speed (satellite), surface temperature, mass flux, and residence time. For the positive krill biomass predictor, as in Equation \eqref{eq:linear-predictor-y}, we considered depth, salinity, and surface temperature. Notably, some drifter-derived products improved the model's performance, as indicated by the WAIC---suggesting that \textit{in situ} and remotely sensed data may provide complementary information to the model.

Figure \ref{fig:aggr_pred} (middle column) shows the predicted krill biomass (log scale) based on the posterior mean. Model estimates were derived from the aggregated data (left column), with predictions based on covariate data from 2020---i.e., \revision{the most recent year for which acoustic krill biomass observations are available}. The corresponding prediction uncertainty, represented by the $2.5^{\text{th}}$ and $97.5^{\text{th}}$ quantiles, is shown in \refmain{Figure SF25} (Supplementary Material). The same figure shows the estimated probabilities (with uncertainty) of observing non-zero krill biomass.

Finally, Figure \ref{fig:aggr_pred} also allows us to visually compare the spatial distribution of krill biomass extrapolated from the acoustic data (middle column) with the corresponding distribution observed in the net haul data (KRILLBASE, right column). Despite the sparse KRILLBASE coverage, we can still identify hotspot areas, particularly in the north-east region around South Georgia Island, which align with findings in the literature \citep{schmidt:2016} and are well captured by our model predictions. However, predictions for the south-west region are more challenging to compare with the net data, meaning that the observed symmetry between the south-west and north-east portions of the map in our predictions (likely driven by environmental factors) should be interpreted with caution. In fact, \cite{brierley:1999} demonstrated substantial differences between the eastern and western parts of the South Georgia shelf, indicating possible fine-scale variability. Taken together, these analyses suggest that our model captures several key spatial patterns observed in the data and may, to a certain extent, reasonably extend these patterns into unsampled regions. Although, as seen in Figure \ref{fig:aggr_pred}, the predicted krill biomass lacks the patchy nature observed in the acoustic and net haul data.

%%%%%%%%%%%%%%%%%%%%%%%%%%%%%%%%%%%%%%%%%%%%%%%%%%
\section{Discussion} \label{sec:discussion}
%%%%%%%%%%%%%%%%%%%%%%%%%%%%%%%%%%%%%%%%%%%%%%%%%%	

In this paper, we presented a statistical framework for modelling krill abundance in the Southern Ocean, with a specific focus on the South Georgia region. By integrating heterogeneous data sources collected at various spatio(-temporal) resolutions and of different types, we tackled key challenges commonly encountered in ecological modelling, such as misaligned datasets and zero-inflated observations. These data sources included acoustic observations of krill biomass, net haul data used for validation, remotely sensed satellite imagery, and drifter-derived covariates. Our approach demonstrated the benefits of combining multiple data sources to enhance both interpretability---evidenced by insights gained from hyperparameter estimation to inform sampling strategies (see Section \ref{ssec:disaggregated})---and predictive accuracy, particularly in the aggregated spatial analysis (see Section \ref{ssec:aggregated}). 

More broadly, integrating remotely sensed data, such as satellite imagery, with timestamped trajectories from drifters offers a powerful approach to modelling marine ecosystems. While satellite imagery provides an overview of environmental conditions, drifter-derived products offer complementary insights into the physical and biological factors influencing the target distribution. These data sources can deliver potentially orthogonal information, even when describing the same phenomenon. In such cases, drifter-derived products can also function as a calibration data source for remotely sensed observations \citep{villejo:2024}. 

The findings of this work may contribute to the development of more effective conservation and management strategies for krill in the Southern Ocean. As highlighted by \cite{warwick:2022}, identifying regions with high krill density enables the determination of areas where krill fishing would have the least ecological impact. 
Moreover, an important consideration when making decisions based on model estimates is the need to account for uncertainty. Misinterpreting or neglecting uncertainty can lead to overconfidence in predictions and potentially harmful management outcomes. By providing credible intervals for key estimates, our framework enables managers to make informed decisions supported by the data while accounting for the inherent variability of ecological systems and the limitations of the modelling process.

% A long-term goal for the krill scientific community is to accurately determine where krill are aggregating across space, enabling fishery management to adapt both spatially and temporally as krill move. Understanding the mechanisms driving krill aggregations and their broader spatio-temporal distribution is essential for achieving this goal. This may require a transition from relying solely on \textit{in situ} acoustic or net haul data, which provide limited temporal and spatial snapshots of krill distributions, to the integration of remote sensing approaches. Remote sensing of krill distributions, similar to the use of satellite chlorophyll data for monitoring phytoplankton, is required to manage fisheries on large scale. Nevertheless, \textit{in situ} data will remain an important resource for validating and refining remote predictions. Such advancements in predicting krill distribution accurately will not only safeguard krill populations but also support the broader ecosystem, including the predators that rely on krill for energy and the species’ significant role in biogeochemical cycles.

Finally, while our work demonstrates the potential of integrating multiple data sources, there are notable limitations. First, relying on acoustic surveys from a smaller region within Subarea 48.3 (green box in Figure \ref{fig:main_map}) makes it challenging to extrapolate predictions beyond the observed window with reasonable uncertainty, particularly for the non-aggregated analysis (Section \ref{ssec:disaggregated}). Second, for the non-aggregated analysis, the assumption of Gaussianity for the underlying random effects in the Hurdle-Gamma model may lead to oversmoothed prediction maps. While this approach reduces computational burden during inference, exploring latent non-Gaussian models (e.g., \cite{cabral:2024}) in future work could offer greater flexibility and more effectively capture abrupt variations in the target ecological variable, even over short distances. \revision{Third, in Section \ref{sec:methods}, we model each covariate with a single linear term; introducing a compact non-linear basis would add flexibility without otherwise modifying the hierarchical framework.} Fourth, although we used net haul data (KRILLBASE) as a validation source for our extrapolated predictions, its spatial and temporal sparsity limits the reliability of conclusions about krill abundance at fine scales. Consequently, comparisons such as those presented in Figure \ref{fig:aggr_pred} should be interpreted with caution, as noted in Section \ref{ssec:aggregated}. Overall, additional extensions could further improve our model. \revision{Given that vessel routes are generally predefined based on prior ecological knowledge---and that sampling locations are determined along these routes---methods that explicitly model the resulting bias, such as that of \cite{amaral:2024}, can be combined with data fusion techniques to improve inference; \citet{zhong:2024} provides one such approach.} Additionally, incorporating other \textit{in situ} data sources, such as profiling floats \citep{roemmich:2009}, could provide complementary oceanographic measurements and refine the modelling of krill abundance. 

\section*{Declarations} 

\subsection*{Conflict of interest} The authors declare that they have no conflict of interest.

\subsection*{Acknowledgments} We thank the Turner-Kirk Trust for supporting this research. We also thank Jeffrey Early for helpful advice in forming drifter-derived spatial products. E. C. was supported by a Natural Environment Research Council (NERC) grant NE/Y004515/1 and a WWF research grant (GB085708). S. F. and data collection for the Western Core Box acoustic survey were funded by the Natural Environment Research Council (NERC).

%\bibliographystyle{cell}
%\bibliography{references}

\end{document}

	% \pagenumbering{gobble}
    \pagenumbering{arabic}
	
	\begin{center}
		{\Large Supplementary Material for ``Navigating Challenges in Spatio-temporal Modelling of Antarctic Krill Abundance: Addressing Zero-inflated Data and Misaligned Covariates''} \\
		\vspace{30pt}
		{\large André Victor Ribeiro Amaral${}^{1, *}$, Adam M. Sykulski${}^{2}$, Sophie Fielding${}^{3}$, Emma Cavan${}^{2}$}\\
	    \vspace{24pt}
        ${}^{1}$University of Southampton. Southampton, UK.\\
	    ${}^{2}$Imperial College London. London, UK.\\
        ${}^{3}$British Antarctic Survey (BAS). Cambridge, UK. \\
		${}^{*}$Corresponding author. E-mail: \texttt{\href{a.v.ribeiro-amaral@soton.ac.uk}{a.v.ribeiro-amaral@soton.ac.uk}}
		\vspace{18pt}
	\end{center}

\newpage
% \pagenumbering{arabic}
\setcounter{footnote}{0} 

%%%%%%%%%%%%%%%%%%%%%%%%%%%%%%%%%%%%%%%%%%%%%%%%%%
\section{Description of Satellite and Drifter-Derived Products} \label{sec:S-products-from-drifters_satellite}
%%%%%%%%%%%%%%%%%%%%%%%%%%%%%%%%%%%%%%%%%%%%%%%%%%

In this section, we provide details on the covariates derived from satellite imagery (\refmain{Section 2.2.1}) and drifter trajectories (\refmain{Section 2.2.2}).

\subsection{Products from Satellite Imagery} \label{ssec:S-products-from-satellite}

We begin by briefly describing the satellite(-derived) products from the Copernicus Marine Service \citep{copernicus:2024} that were used in our analyses.

\begin{itemize}
\item ``Chlorophyll:'' mass concentration of chlorophyll \textit{a} in seawater ($\text{mg}\cdot\text{m}^{-3}$), averaged over 75 depth levels ranging from 0.51 metres to 5902.06 metres. It is a product from the ``Global Ocean Biogeochemistry Hindcast'' (\href{https://doi.org/10.48670/moi-00019}{\texttt{10.48670/moi-00019}}).
\item ``Potential temperature:'' the temperature a parcel of seawater would have if moved adiabatically (without any exchange of heat with its surroundings) to the sea surface (${}^{\circ}\text{C}$), averaged over 49 depth levels ranging from 0.49 metres to 5727.92 metres. It is a product from ``Global Ocean Physics Reanalysis'' (\href{https://doi.org/10.48670/moi-00021}{\texttt{10.48670/moi-00021}}).
\item ``Salinity:'' sea-water salinity (dimensionless, reported as $10^{-3}$ or \textperthousand), averaged over 49 depth levels ranging from 0.49 metres to 5727.92 metres. It is a product from ``Global Ocean Physics Reanalysis''  (\href{https://doi.org/10.48670/moi-00021}{\texttt{10.48670/moi-00021}}).
\item ``Speed (satellite):'' computed as $\sqrt{u^2 + v^2}$, where $u$ and $v$ are the eastward and northward seawater velocities $(\text{m} \cdot \text{s}^{-1})$, respectively. Values were averaged over 49 depth levels ranging from 0.49 metres to 5727.92 metres. \hspace{2pt}It is a product from ``Global Ocean Physics Reanalysis''  (\href{https://doi.org/10.48670/moi-00021}{\texttt{10.48670/moi-00021}}).
\item ``Surface temperature:'' sea surface temperature ($\text{K}$). It is a product from ``Global High Resolution ODYSSEA Sea Surface Temperature Multi-sensor L3 Observations'' (\href{https://doi.org/10.48670/mds-00329}{\texttt{10.48670/mds-00329}}).
\end{itemize}

In addition to these products, we also used bathymetry data (depth) from the National Centers for Environmental Information (NOAA;  \href{https://doi.org/10.25921/fd45-gt74}{\texttt{10.25921/fd45-gt74}}) and slope, which was computed from the bathymetry.

\subsection{Products from Drifter Trajectories} \label{ssec:S-products-from-drifters}

Next, we outline the procedure that is followed to compute the drifter-derived products described in \refmain{Section 3.2}.

\begin{figure}[!ht]
	\centering
	\includegraphics[width = 0.6\textwidth]{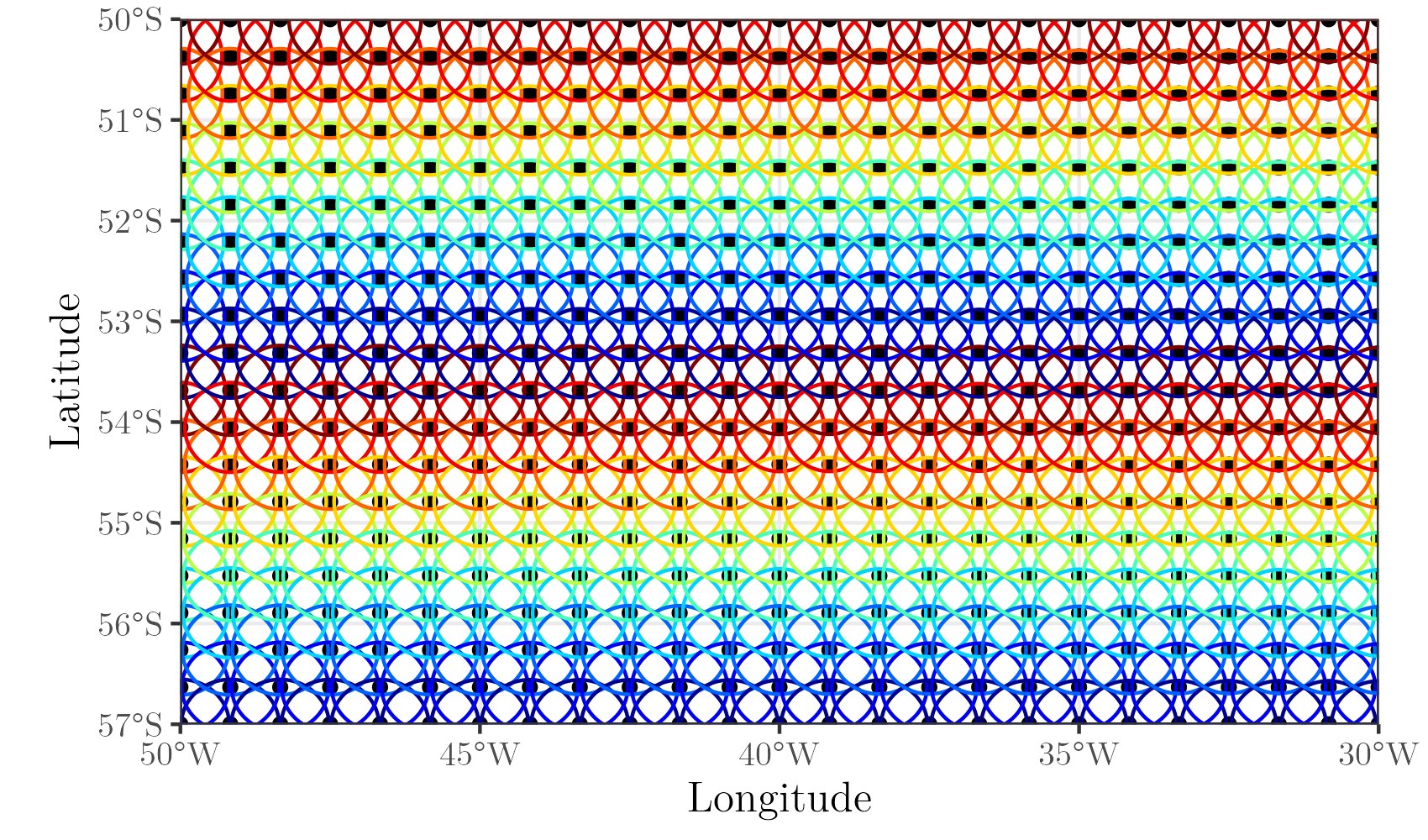}
    \caption{Spatial windows (circles of radius 50 km) used to compute ``residence time'' and ``mass flux.'' The colours are only used to ease the visualisation.}
	\label{fig:circles_drifters}
\end{figure}

\begin{itemize}
\item ``Surface speed (drifters):'' as stated in \refmain{Table 1}, this quantity was computed based on drifter trajectories observed during the months of December, January, and February, so that it is comparable with the speed estimates obtained from the satellite imagery.
\item ``Expected frequency:'' it was computed over rolling temporal windows of 121 hours in length, with a 60-hour overlap between consecutive segments. The location of a segment is defined by its mid-point.
\item ``Residence time'' and ``Mass flux:'' in both cases, the spatial windows were defined by circles of radius 50 km, with centres at $20 \times 25$ points distributed equidistantly over the study area. The location of a circle is defined by its centre point. Figure \ref{fig:circles_drifters} shows these circles. To avoid border effects, the estimates in the circles at the edges were corrected by a multiplicative factor given by the ratio between the area of a full circle and the area of the clipped circle.
\item ``Density of drifters:'' the computation is straightforward and corresponds to the total number of hours spent by drifters in a certain spatial area, specifically a $0.25^{\circ}$ longitude by $0.25^{\circ}$ latitude cell, divided by the area of this cell.
\end{itemize}

\subsubsection{Interpolation} \label{sssec:S-drifter-interpolation}

After computing all drifter statistics (except for the ``density of drifters'') as described in \refmain{Section 3.2}, we interpolate these newly generated data points across the entire study area shown in \refmain{Figure 1} (South Georgia, Subarea 48.3) using the following approach. 

Let $y(s) = (y(s_1), \cdots, y(s_n))$ represent the drifter product observed at locations $(s_1, \cdots, s_n) \subset \mathcal{X} \subset \mathbb{R}^2$, where $\mathcal{X}$ is the spatial domain. To spatially interpolate these observations across $\mathcal{X}$, we model $y(s)$ as follows
\begin{align*}
    y(s) &= \beta_0 + \phi(s) + \epsilon(s), \text{ s.t. } \epsilon(s) \overset{\text{i.i.d.}}{\sim} \text{Normal}(0, \sigma^2_\epsilon) \\
    \phi(s) &\sim \text{Gaussian Process}(0, r_{\phi}(h; \theta)) \\
    (\beta_0, \sigma^2_{\epsilon}, \theta) &\sim \text{priors},
\end{align*}
where $r_{\phi}(h; \theta)$ is the Matérn is a Matérn covariance kernel with parameters $\theta = (\nu, \kappa, \sigma_{\phi})$. The model is fitted with \texttt{R-INLA} \citep{lindgren:2015}; technical details of the INLA implementation are given in \refmain{Section 3.1.2}.

We also note that, although the true drifter-derived surfaces may exhibit local anisotropy and non-stationarity, we adopt a stationary Gaussian process model as a pragmatic compromise between realism and computational efficiency. However, this choice can be replaced, if needed, with a more sophisticated interpolation approach, such as the one described in \cite{lodise:2020}.

%%%%%%%%%%%%%%%%%%%%%%%%%%%%%%%%%%%%%%%%%%%%%%%%%%
\section{Spatial Modelling} \label{sec:S-spatial-equi}
%%%%%%%%%%%%%%%%%%%%%%%%%%%%%%%%%%%%%%%%%%%%%%%%%%

In \refmain{Section 3.2}, we discussed how the aggregated spatial-only analysis may be seen as an approximation to the time-averaged mean surface of the full spatio-temporal model. To illustrate this equivalence, we present a simulation study.

\begin{figure}[!ht]
	\centering
	\includegraphics[width = 1\textwidth]{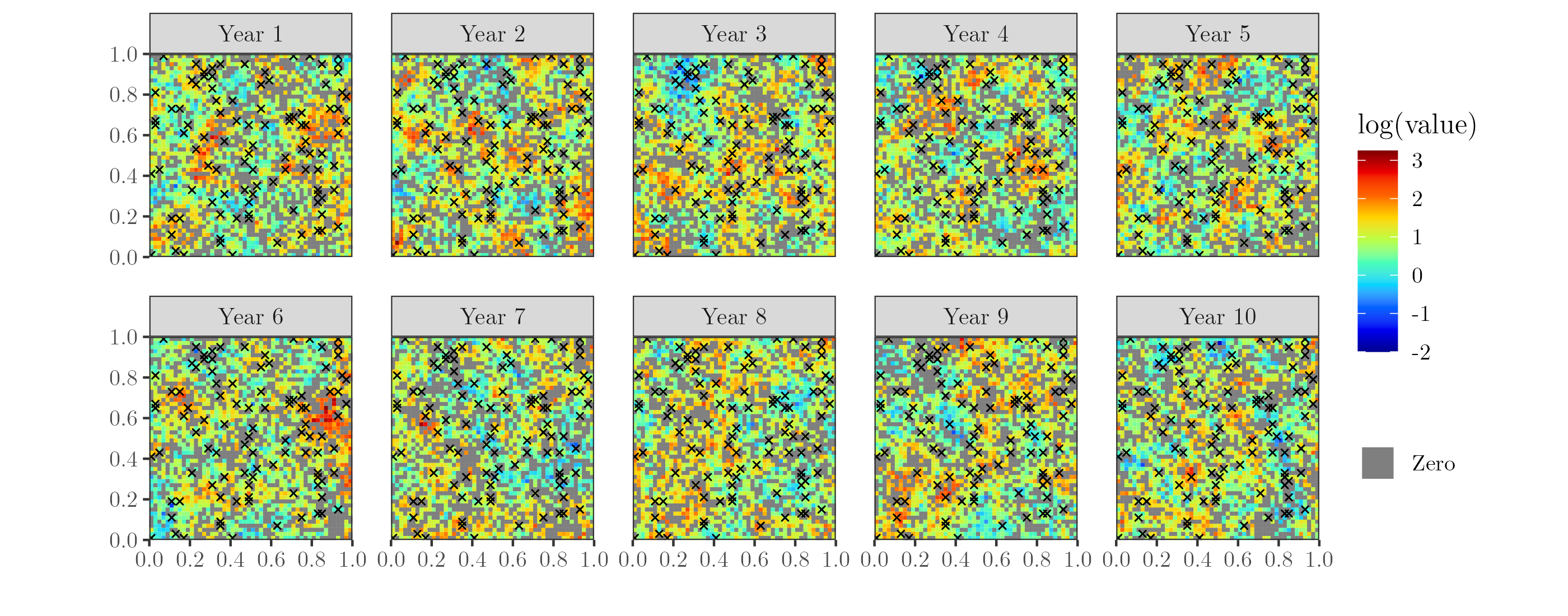}
    \caption{Simulation from a spatio-temporal Hurdle-Gamma model defined on $[0, 1] \times [0, 1]$. Observations $z_{it} \cdot y_{it}$, for $i = 1, \cdots, n$ and $t = 1, \cdots, 10$, are indicated by crosses ($\times$).}
	\label{fig:spatial_sim}
\end{figure}

We simulate data from the model in \refmain{Section 3.1} on the unit square (i.e., $\mathcal{X} = [0, 1] \times [0, 1]$), for $t = 1, \cdots, T$, where $T = 10$ years. In particular, $z_{it} \sim \text{Bernoulli}(\pi_{it})$ and $y_{it} \sim \text{Gamma}(a_{it}, b_{it})$, such that $\text{logit}(\pi_{it}) = \beta_0^z + \beta_1^z \texttt{cov}_{it} + \psi_{it}$ and $\log(\mu_{it}) = \beta_0^y + \beta_1^y \texttt{cov}_{it} + \gamma \cdot \psi_{it} + \xi_{it}$, where $\mathbb{E}(y_{it}) = \mu_{it} = a_{it}/b_{it}$. The spatio-temporal random effects ($\psi_{it}$ and $\xi_{it}$) are defined as in \refmain{Equation (3)} in the main paper, and the true (and estimated) parameters are shown in Table \ref{tab:S-est_coeff_spatial}. Furthermore, $\texttt{cov}_{it} \overset{\text{i.i.d. in $t$}}{\sim}\text{GP}_{\text{Matérn}}(0, \sigma^2 = 1, \nu = 1, \rho = 0.2)$. Figure \ref{fig:spatial_sim} shows the simulated data $z_{it} \cdot y_{it}$, such that $i = 1, \cdots, 100$ and $t = 1, \cdots, 10$.

\begin{table}[!ht]
	\caption{True and estimated parameters for the a spatio-temporal Hurdle-Gamma model. The parameter estimates are based on the mean and standard deviation of the corresponding posterior distribution. Note that $\nu$ is not estimated but fixed at 1 in all cases, both in simulation and inference. For the Gamma distribution, we set $a_{it} = k$ and $b_{it} = k/\mu_{it}$.}
    %\resizebox{\textwidth}{!}{%
    \centering
    \begin{tabular}{ c | c | c | c | c | c }
    Parameter & True & Estimated & Parameter & True & Estimated \\ \hline
    $\beta_0^z$ & \phantom{-1}0.5 & \phantom{-}0.466 (0.105) & $\rho_{\phi, \psi}$   & 0.25 & 0.180 (0.070) \\
    $\beta_0^y$ & \phantom{-1}1.0 & \phantom{-}0.976 (0.030) & $\sigma_{\phi, \psi}$ & 0.44 & 0.574 (0.156) \\
    $\beta_1^z$ & \phantom{-1}0.3 & \phantom{-}0.221 (0.069) & $\alpha_{\psi}$       & 0.40 & 0.782 (0.124) \\
    $\beta_1^z$ & \phantom{1}-0.4 & \phantom{}-0.409 (0.016) & $\rho_{\phi, \xi}$    & 0.20 & 0.306 (0.106) \\
    $k$         & \phantom{-}10.0 & \phantom{-}9.008 (0.866) & $\sigma_{\phi, \xi}$  & 0.14 & 0.176 (0.032) \\
    $\gamma$    & \phantom{-1}0.3 & \phantom{-}0.250 (0.086) & $\alpha_{\xi}$        & 0.20 & 0.044 (0.247)
    \end{tabular}
    %}
	\label{tab:S-est_coeff_spatial}
\end{table}

\begin{figure}[!ht]
	\centering
	\includegraphics[width = 0.7\textwidth]{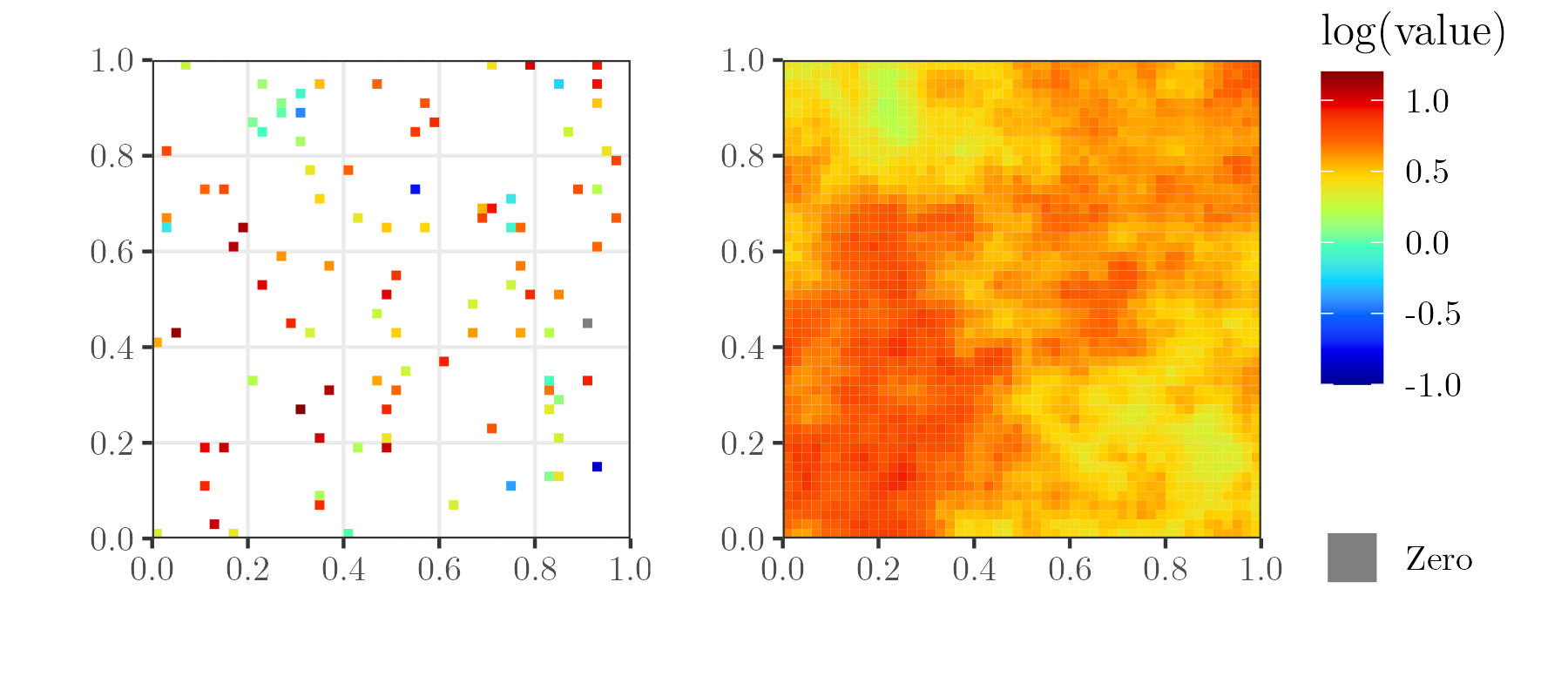} \vspace{-10pt}
    \caption{Left: Temporally aggregated (10-year) simulated data. Right: Fitted positive-state component ($\hat{\mu}_i$), based on the posterior mean from the spatial-only model.}
	\label{fig:spatial_fit_sim}
\end{figure}

Inference was performed using \texttt{R-INLA} \citep{lindgren:2015}, as described in \refmain{Section 3.1.2}. The estimated parameters, reported in Table~\ref{tab:S-est_coeff_spatial}, were adequately recovered under the correctly specified spatio-temporal model. However, our primary interest lies in comparing the predictions from this approach with those from a spatial-only model fitted using a temporally aggregated version of the data shown in Figure \ref{fig:spatial_sim}. To this end, we fit a spatial model as described in \refmain{Section 3.1.1}, where both the covariate and response observations were aggregated by averaging over the 10-year period. Figure \ref{fig:spatial_fit_sim} shows the aggregated data at the observed locations alongside the corresponding predictions, while Table \ref{tab:S-spatial-vs-st} compares the two fits in terms of the time-averaged means
\begin{align*}
    \hat{\theta}_i^{\text{spatio-temporal}} = \frac{1}{T}\sum_{t = 1}^{T} (\hat{\pi}_{it} \cdot \hat{\mu}_{it})~~~\text{vs.}~~~~\hat{\theta}_i^{\text{spatial}} = (\hat{\pi}_{i} \cdot \hat{\mu}_{i}),
\end{align*}
where $\hat{\pi}_{it}$ and $\hat{\mu}_{it}$ are the posterior means of the presence probability and the positive-state component in year $t$, and $\hat{\pi}_i$ and $\hat{\mu}_i$ are the corresponding posterior means from the spatial-only model.

\begin{table}[!ht]
	\caption{Comparison between the spatial and spatio-temporal models with respect to the time-averaged means at 100 observed locations. For each model, 500 posterior samples were drawn per location, and the posterior mean and 95\% equal-tail credible interval were computed at each site. Model performance was evaluated by comparing the posterior summaries to the corresponding true values using the following metrics: RMSE (root mean squared error), RMSPE (root mean squared percentage error), MAE (mean absolute error), MAPE (mean absolute percentage error), $\text{Width}_{95\%}$ (mean 95\% equal-tail interval width), and $\text{Coverage}_{95\%}$ (empirical 95\% coverage). All metrics were averaged across the 100 locations, except for \hspace{1pt}$\text{Coverage}_{95\%}$, which represents the proportion of locations covered.    
    %$\text{RMSE} = \sqrt{n^{-1}\sum_{i=1}^n(\hat{\theta}_i - \theta_i)^2}$ (root mean squared error), $\text{RMSPE} = \sqrt{n^{-1}\sum_{i=1}^n\left((\hat{\theta}_i - \theta_i)/{\theta_i}\right)^2}$ (root mean squared percentage error), $\text{MAE} = n^{-1}\sum_{i = 1}^n|\hat{\theta}_i - \theta_i|$ (mean absolute error), $\text{MAPE} = n^{-1}\sum_{i = 1}^n\left|(\hat{\theta}_i - \theta_i)/{\theta_i}\right|$ (mean absolute percentage error), and $\text{Width}_{95} = n^{-1}\sum_{i = 1}^n(\hat{\theta}_{i, 0.975} - \hat{\theta}_{i, 0.025})$ (mean 95\% equal-tail interval width), and $\text{Coverage}_{95} = n^{-1} \sum_{i = 1}^n \mathds{1}\left\{\theta_i \in \left[ \hat{\theta}_{i, 0.025}, \hat{\theta}_{i, 0.975} \right]\right\}$ (mean 95\% coverage). Here, $\theta_i$ denotes the time-averaged mean for cell $i$ in the simulated \textit{truth}, and $\hat{\theta}_i$ is the corresponding posterior mean.
    }
    %\resizebox{\textwidth}{!}{%
    \centering
    \begin{tabular}{ c | c | c | c | c | c | c }
    Model & RMSE & RMSPE & MAE & MAPE & $\text{Width}_{95\%}$ & $\text{Coverage}_{95\%}$\\ \hline
    Spatial\phantom{-emporal} & 0.264 & 0.159 & 0.212 & 0.124 & 0.948 & 0.900 \\
    Spatio-temporal           & 0.262 & 0.150 & 0.207 & 0.116 & 1.090 & 0.970
    \end{tabular}
    %}
	\label{tab:S-spatial-vs-st}
\end{table}

Table \ref{tab:S-spatial-vs-st} shows that the spatial and spatio-temporal models achieve similar predictive accuracy for the time-averaged means. This suggests that, under the assumptions outlined in \refmain{Section 3.1.1}, the bias introduced by using a temporally aggregated spatial model may be negligible. However, in this setting, the spatio-temporal model performs consistently better and achieves superior 95\% coverage. These results support the equivalence between the two approaches discussed in the same section, while also quantifying the modest loss of efficiency due to aggregation in this specific scenario.

%%%%%%%%%%%%%%%%%%%%%%%%%%%%%%%%%%%%%%%%%%%%%%%%%%
\section{Additional Results} \label{sec:S-additional_results}
%%%%%%%%%%%%%%%%%%%%%%%%%%%%%%%%%%%%%%%%%%%%%%%%%%

In this section, we present additional analyses to complement the results from \refmain{Section 4}, including details on both disaggregated spatio-temporal modelling and aggregated spatial modelling.

\subsection{Disaggregated Spatio-temporal Modelling} \label{ssec:S-disagg_sup}

In \refmain{Section 4.1}, we implemented a Hurdle-Gamma model (as stated in \refmain{Section 3.1}) with linear predictors specified as follows
\begin{align} \label{eq:lp_z}
    \text{logit}(\pi_{it}) = ~
    &\beta_0^z + \beta_1^z \texttt{month} + 
    \beta_2^z \texttt{depth}_{i} + 
    \beta_3^z \texttt{slope}_{i} +\\ 
    &\beta_4^z \texttt{chlor}_{it} +
    \beta_5^z \texttt{pot\_temp}_{it} + 
    \beta_6^z \texttt{salinity}_{it} + 
    \beta_7^z \texttt{speed\_sat}_{it} +
    \beta_8^z \texttt{surf\_temp}_{it} +\nonumber\\
    &\beta_9^z \texttt{speed\_drif}_{it} +
    \beta_{10}^z \texttt{expect\_freq}_{it} +
    \beta_{11}^z \texttt{res\_time}_{it} +
    \beta_{12}^z \texttt{mass\_flux}_{it} +
    \beta_{13}^z \texttt{density\_drif}_{it} +\nonumber\\ 
    &\psi_{it}, \nonumber
\end{align}
and
\begin{align} \label{eq:lp_y}
    \log(\mu_{it}) = ~
    &\beta_0^y + \beta_1^z \texttt{month} + 
    \beta_2^y \texttt{depth}_{i} + 
    \beta_3^y \texttt{slope}_{i} +\\ 
    &\beta_4^y \texttt{chlor}_{it} +
    \beta_5^y \texttt{pot\_temp}_{it} + 
    \beta_6^y \texttt{salinity}_{it} + 
    \beta_7^y \texttt{speed\_sat}_{it} +
    \beta_8^y \texttt{surf\_temp}_{it} +\nonumber\\
    &\beta_9^y \texttt{speed\_drif}_{it} +
    \beta_{10}^y \texttt{expect\_freq}_{it} +
    \beta_{11}^y \texttt{res\_time}_{it} +
    \beta_{12}^y \texttt{mass\_flux}_{it} +
    \beta_{13}^y \texttt{density\_drif}_{it} +\nonumber\\ 
    &\gamma \cdot \psi_{it} + \xi_{it}, \nonumber
\end{align}
where $\gamma$ is a ``copy'' factor, and $\psi_{it}$ (similarly, $\xi_{it}$) is spatio-temporal random effect, such that
\begin{align*} 
    \psi_{it} &= \alpha_{\psi} \psi_{i, (t - 1)} + \phi_{it}, ~~ t = 1997, \cdots, 2020, 
\end{align*}
where $|\alpha_{\phi}| <1$, $\psi_{i1} \sim \text{Normal}(0, \sigma^2_{\phi, \psi}/(1 - \alpha_{\phi}^2))$, and $\phi_{it}$ is a temporally independent but spatially dependent Gaussian process at each time point, with a Matérn covariance structure with range $\rho_{\phi, \psi}$ and marginal variance $\sigma_{\phi, \psi}^2$.

In Equations \eqref{eq:lp_z} and \eqref{eq:lp_y}, all covariates are defined as in \refmain{Table 1}, and \texttt{month} refers to the month in which the krill acoustic data were collected each year, with possible values of December, January, and February, mapped to $-1$, $0$, and $1$, respectively (where December data are considered observations for the following year). For numerical stability, all covariates were scaled to have a mean of zero and a variance of one when fitting the model. The estimated coefficients are shown in Table \ref{tab:S-est_coeff_disagg}.

\begin{table}[!ht]
	\caption{Estimated parameters (with standard deviation and a 95\% equal-tail credible interval) for the spatio-temporal model fitted for the disaggregated data.}
    \resizebox{\textwidth}{!}{%
    \centering
    \begin{tabular}{ c | c | c | c | c | c | c | c | c | c | c | c }
    Parameter & Mean & SD & 95\% equal-tail CI & Parameter & Mean & SD & 95\% equal-tail CI & Parameter & Mean & SD & 95\% equal-tail CI \\ \hline
    $\beta_{0}^z$  & $\phantom{-}0.928$ & $0.284$ & $(\phantom{-}0.372; \phantom{-}1.485)$ & $\beta_{0}^y$  & $\phantom{-}2.397$ & $0.063$ & $(\phantom{-}2.273; \phantom{-}2.521)$ & $\rho_{\phi,\psi}$   & $37.663$ & $1.965$ & $(34.115; 41.843)$ \\
    $\beta_{1}^z$  & $-0.104$           & $0.325$ & $(-0.741; \phantom{-}0.534)$           & $\beta_{1}^y$  & $-0.287$           & $0.082$ & $(-0.448; -0.126)$                     & $\sigma_{\phi,\psi}$ & $\phantom{0}2.974$ & $0.106$ & $(\phantom{0}2.774; \phantom{0}3.193)$ \\
    $\beta_{2}^z$  & $\phantom{-}0.176$ & $0.170$ & $(-0.156; \phantom{-}0.508)$           & $\beta_{2}^y$  & $-0.142$ & $0.081$ & $(-0.016; -0.301)$                     & $\alpha_{\psi}$      & $\phantom{0}0.122$ & $0.035$ & $(\phantom{0}0.054; \phantom{0}0.192)$ \\
    $\beta_{3}^z$  & $\phantom{-}0.009$ & $0.033$ & $(-0.056; \phantom{-}0.074)$           & $\beta_{3}^y$  & $\phantom{-}0.035$ & $0.019$ & $(-0.001; \phantom{-}0.072)$           & $\rho_{\phi,\xi}$    & $\phantom{0}6.954$ & $0.547$ & $(\phantom{0}6.001; \phantom{0}8.151)$ \\
    $\beta_{4}^z$  & $-0.070$           & $0.098$ & $(-0.263; \phantom{-}0.123)$           & $\beta_{4}^y$  & $\phantom{-}0.009$ & $0.057$ & $(-0.102; \phantom{-}0.120)$           & $\sigma_{\phi,\xi}$  & $\phantom{0}2.295$ & $0.086$ & $(\phantom{0}2.122; \phantom{0}2.462)$ \\
    $\beta_{5}^z$  & $\phantom{-}0.209$ & $0.162$ & $(-0.108; \phantom{-}0.527)$           & $\beta_{5}^y$  & $-0.308$           & $0.061$ & $(-0.429; -0.188)$                     & $\alpha_{\xi}$       & $\phantom{0}0.064$ & $0.032$ & $(\phantom{0}0.000; \phantom{0}0.128)$ \\
    $\beta_{6}^z$  & $-0.037$           & $0.158$ & $(-0.346; \phantom{-}0.273)$           & $\beta_{6}^y$  & $-0.279$           & $0.076$ & $(-0.428; -0.130)$                     & $\gamma$             & $\phantom{0}0.017$ & $0.018$ & $(-0.017; \phantom{0}0.053)$ \\
    $\beta_{7}^z$  & $\phantom{-}0.172$ & $0.115$ & $(-0.052; \phantom{-}0.397)$           & $\beta_{7}^y$  & $\phantom{-}0.021$ & $0.052$ & $(-0.081; \phantom{-}0.123)$           &&&& \\
    $\beta_{8}^z$  & $-0.131$           & $0.096$ & $(-0.319; \phantom{-}0.056)$           & $\beta_{8}^y$  & $\phantom{-}0.068$ & $0.050$ & $(-0.030; \phantom{-}0.165)$           &&&& \\
    $\beta_{9}^z$  & $-0.019$           & $0.064$ & $(-0.145; \phantom{-}0.107)$           & $\beta_{9}^y$  & $-0.059$           & $0.042$ & $(-0.142; \phantom{-}0.024)$           &&&& \\
    $\beta_{10}^z$ & $\phantom{-}0.280$ & $0.185$ & $(-0.082; \phantom{-}0.642)$           & $\beta_{10}^y$ & $-0.020$           & $0.080$ & $(-0.177; \phantom{-}0.138)$           &&&& \\
    $\beta_{11}^z$ & $\phantom{-}0.152$ & $0.223$ & $(-0.284; \phantom{-}0.589)$           & $\beta_{11}^y$ & $\phantom{-}0.102$ & $0.094$ & $(-0.082; \phantom{-}0.286)$           &&&& \\
    $\beta_{12}^z$ & $-0.106$           & $0.211$ & $(-0.520; \phantom{-}0.307)$           & $\beta_{12}^y$ & $-0.050$           & $0.083$ & $(-0.213; \phantom{-}0.112)$           &&&& \\
    $\beta_{13}^z$ & $\phantom{-}0.000$ & $0.064$ & $(-0.125; \phantom{-}0.126)$           & $\beta_{13}^y$ & $-0.055$           & $0.033$ & $(-0.121; \phantom{-}0.010)$           &&&& \\
    
    \end{tabular}
    }
	\label{tab:S-est_coeff_disagg}
\end{table}

Table \ref{tab:S-est_coeff_disagg} shows that several fixed-effect 95\% credible intervals overlap zero, providing little evidence that those covariates influence krill biomass. In \refmain{Section 4.2} we exploit the lower computational cost of the spatial-only model to perform stepwise forward selection, resulting in a more parsimonious specification.

\begin{figure}[!ht]
	\centering
	\includegraphics[width = 0.9\textwidth]{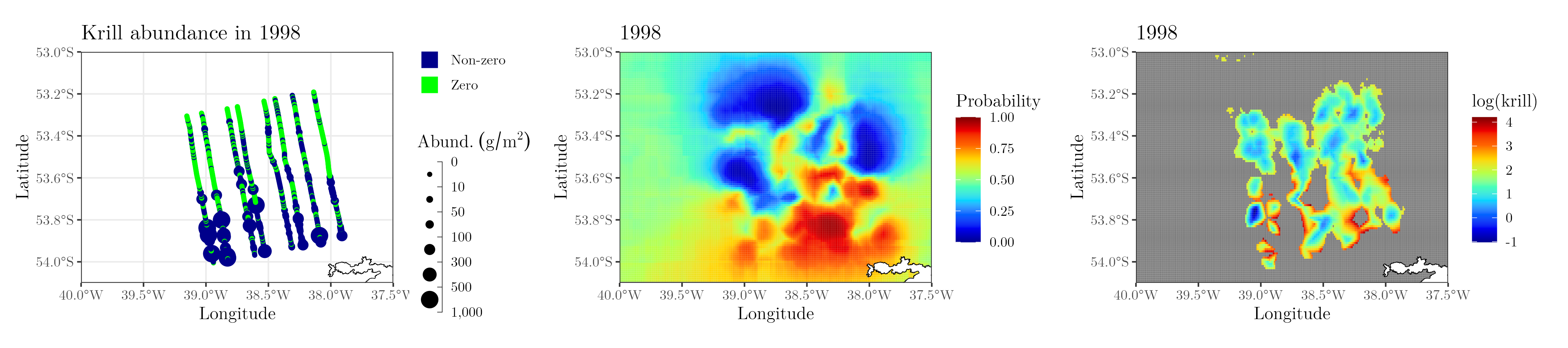}
    \caption{Year 1998. Left: observed acoustic krill biomass data. Middle: estimated probability of non-zeros. Right: predicted krill biomass ($\text{g}/\text{m}^2$), with predictions having a standard deviation greater than 3 (on the log scale) being masked out. The two rightmost plots are based on the mean of the corresponding predictive distributions.}
	\label{fig:all_pred_original_data_1998}
\end{figure}

\begin{figure}[!ht]
	\centering
	\includegraphics[width = 0.9\textwidth]{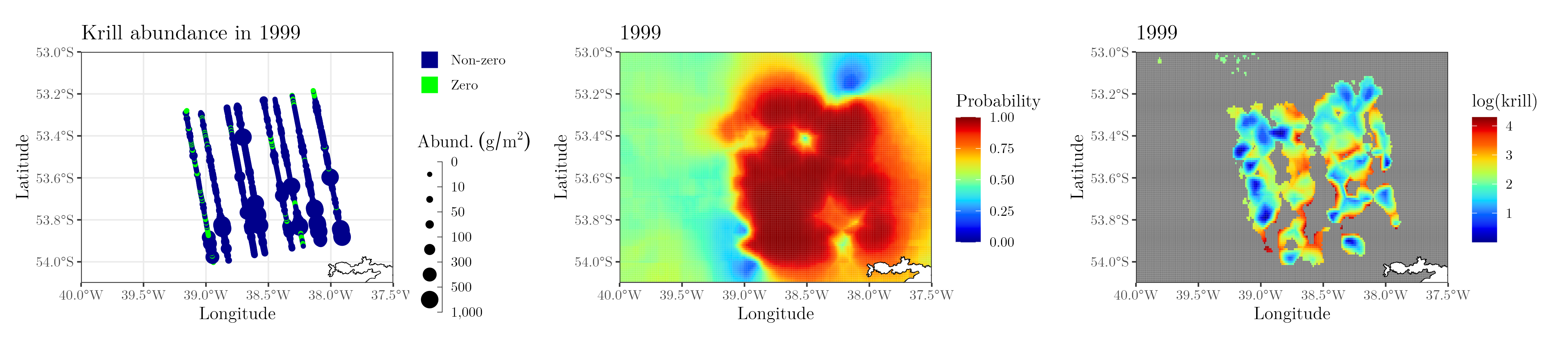}
    \caption{Year 1999. Left: observed acoustic krill biomass data. Middle: estimated probability of non-zeros. Right: predicted krill biomass ($\text{g}/\text{m}^2$), with predictions having a standard deviation greater than 3 (on the log scale) being masked out. The two rightmost plots are based on the mean of the corresponding predictive distributions.}
	\label{fig:all_pred_original_data_1999}
\end{figure}

\begin{figure}[!ht]
	\centering
	\includegraphics[width = 0.9\textwidth]{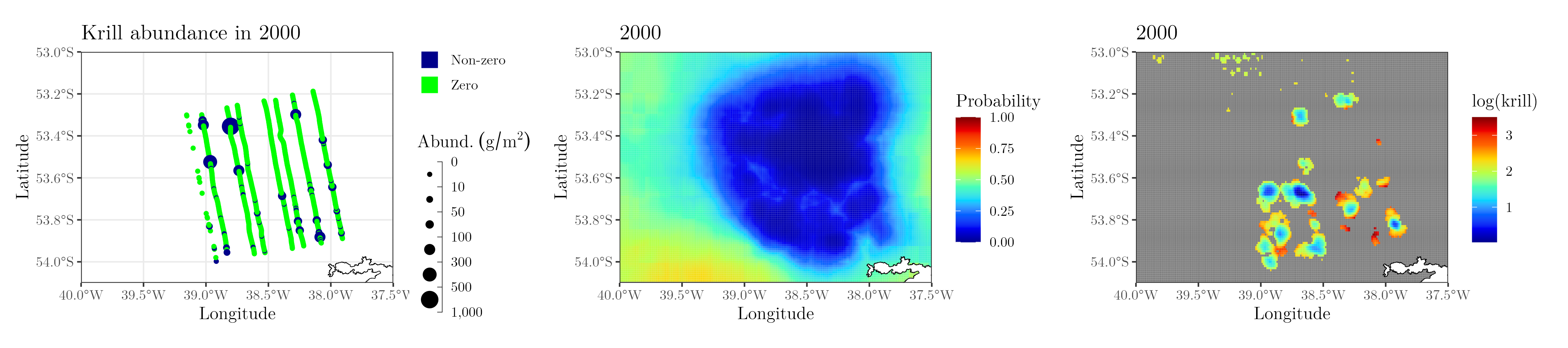}
    \caption{Year 2000. Left: observed acoustic krill biomass data. Middle: estimated probability of non-zeros. Right: predicted krill biomass ($\text{g}/\text{m}^2$), with predictions having a standard deviation greater than 3 (on the log scale) being masked out. The two rightmost plots are based on the mean of the corresponding predictive distributions.}
	\label{fig:all_pred_original_data_2000}
\end{figure}

\begin{figure}[!ht]
	\centering
	\includegraphics[width = 0.9\textwidth]{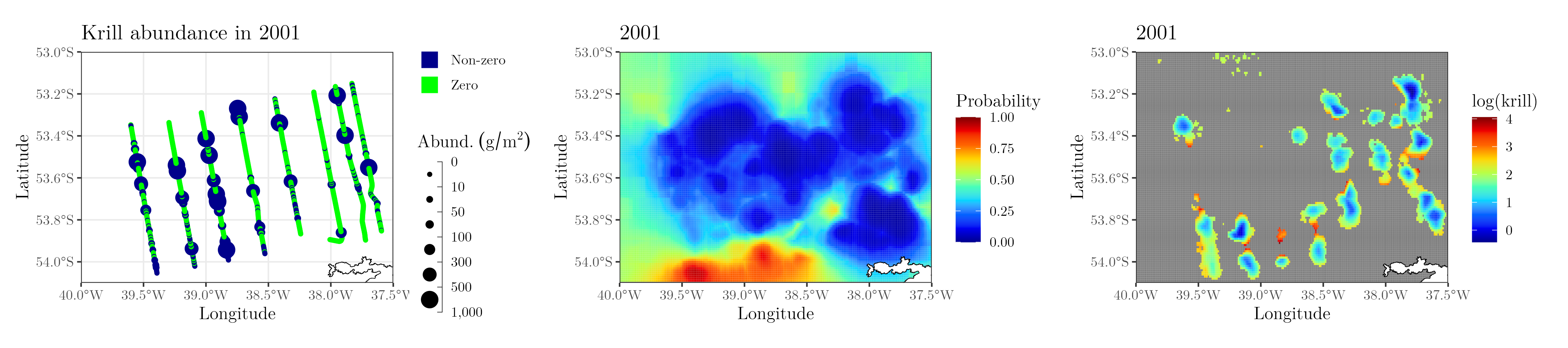}
    \caption{Year 2001. Left: observed acoustic krill biomass data. Middle: estimated probability of non-zeros. Right: predicted krill biomass ($\text{g}/\text{m}^2$), with predictions having a standard deviation greater than 3 (on the log scale) being masked out. The two rightmost plots are based on the mean of the corresponding predictive distributions.}
	\label{fig:all_pred_original_data_2001}
\end{figure}

\begin{figure}[!ht]
	\centering
	\includegraphics[width = 0.9\textwidth]{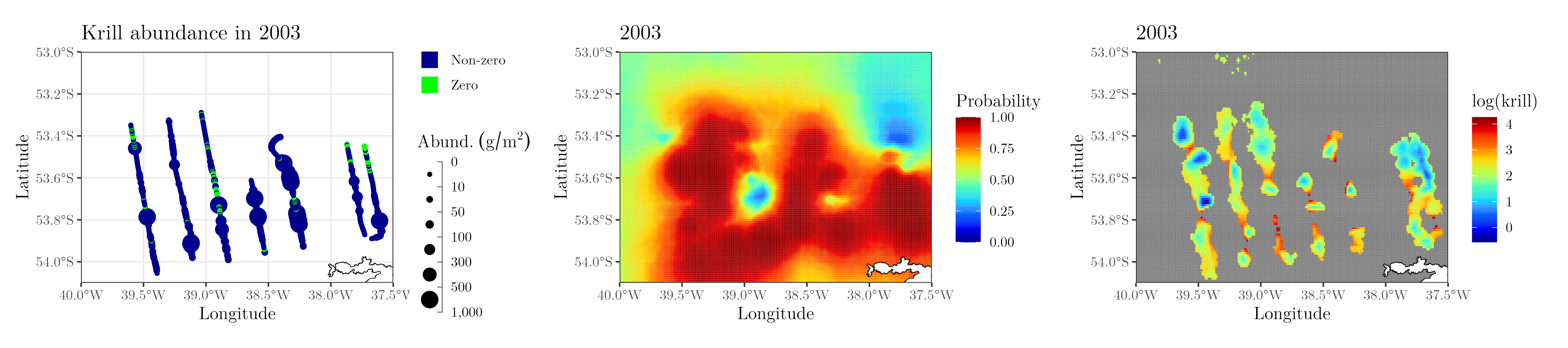}
    \caption{Year 2003. Left: observed acoustic krill biomass data. Middle: estimated probability of non-zeros. Right: predicted krill biomass ($\text{g}/\text{m}^2$), with predictions having a standard deviation greater than 3 (on the log scale) being masked out. The two rightmost plots are based on the mean of the corresponding predictive distributions.}
	\label{fig:all_pred_original_data_2003}
\end{figure}

\begin{figure}[!ht]
	\centering
	\includegraphics[width = 0.9\textwidth]{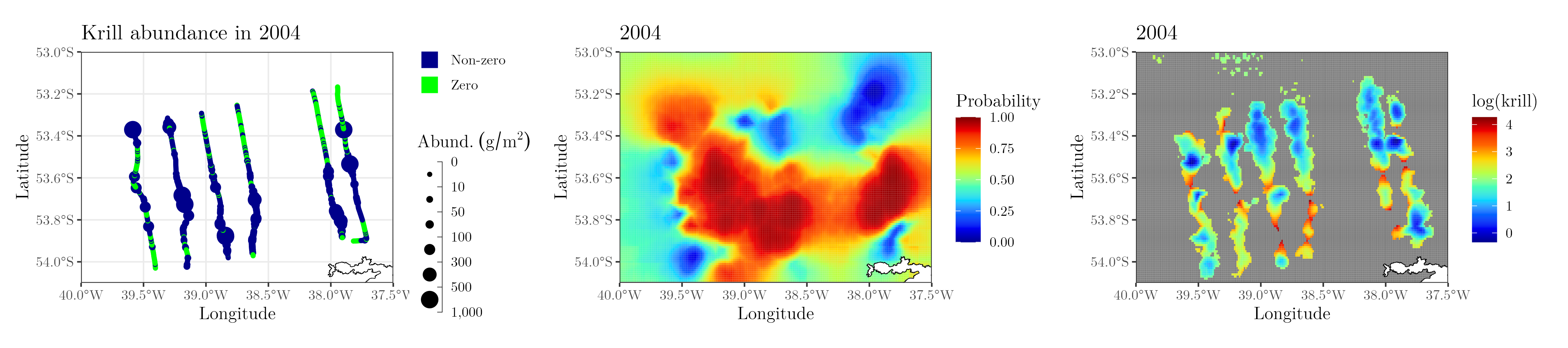}
    \caption{Year 2004. Left: observed acoustic krill biomass data. Middle: estimated probability of non-zeros. Right: predicted krill biomass ($\text{g}/\text{m}^2$), with predictions having a standard deviation greater than 3 (on the log scale) being masked out. The two rightmost plots are based on the mean of the corresponding predictive distributions.}
	\label{fig:all_pred_original_data_2004}
\end{figure}

\begin{figure}[!ht]
	\centering
	\includegraphics[width = 0.9\textwidth]{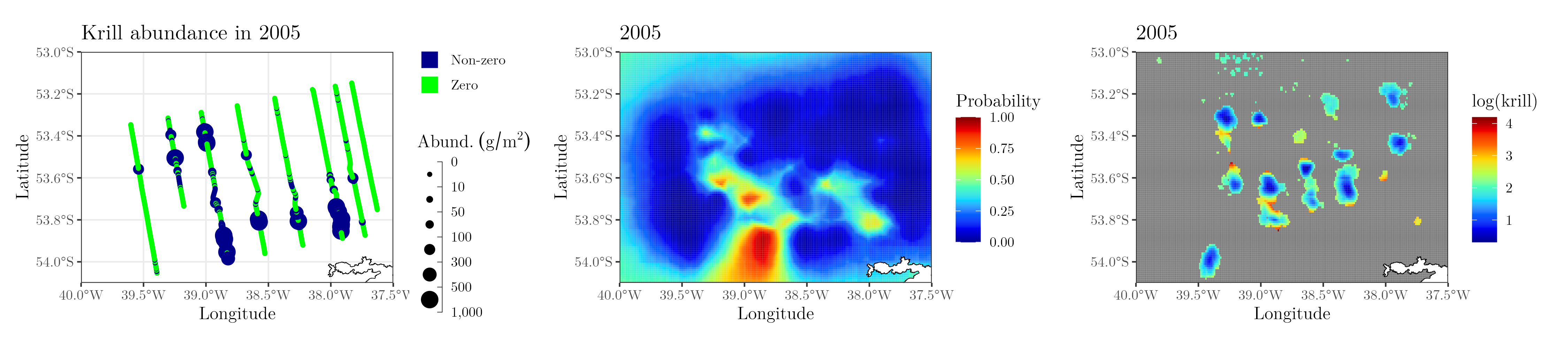}
    \caption{Year 2005. Left: observed acoustic krill biomass data. Middle: estimated probability of non-zeros. Right: predicted krill biomass ($\text{g}/\text{m}^2$), with predictions having a standard deviation greater than 3 (on the log scale) being masked out. The two rightmost plots are based on the mean of the corresponding predictive distributions.}
	\label{fig:all_pred_original_data_2005}
\end{figure}

\begin{figure}[!ht]
	\centering
	\includegraphics[width = 0.9\textwidth]{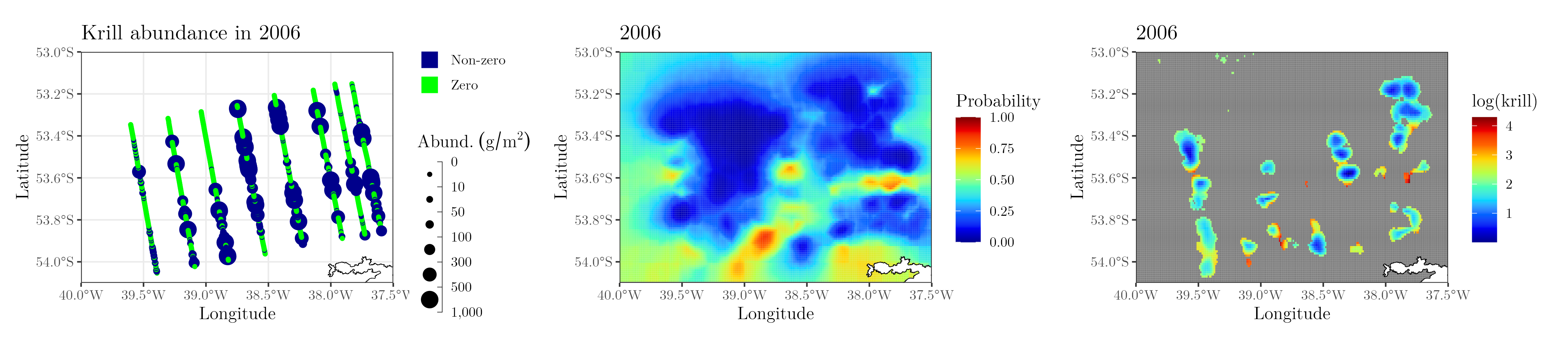}
    \caption{Year 2006. Left: observed acoustic krill biomass data. Middle: estimated probability of non-zeros. Right: predicted krill biomass ($\text{g}/\text{m}^2$), with predictions having a standard deviation greater than 3 (on the log scale) being masked out. The two rightmost plots are based on the mean of the corresponding predictive distributions.}
	\label{fig:all_pred_original_data_2006}
\end{figure}

\begin{figure}[!ht]
	\centering
	\includegraphics[width = 0.9\textwidth]{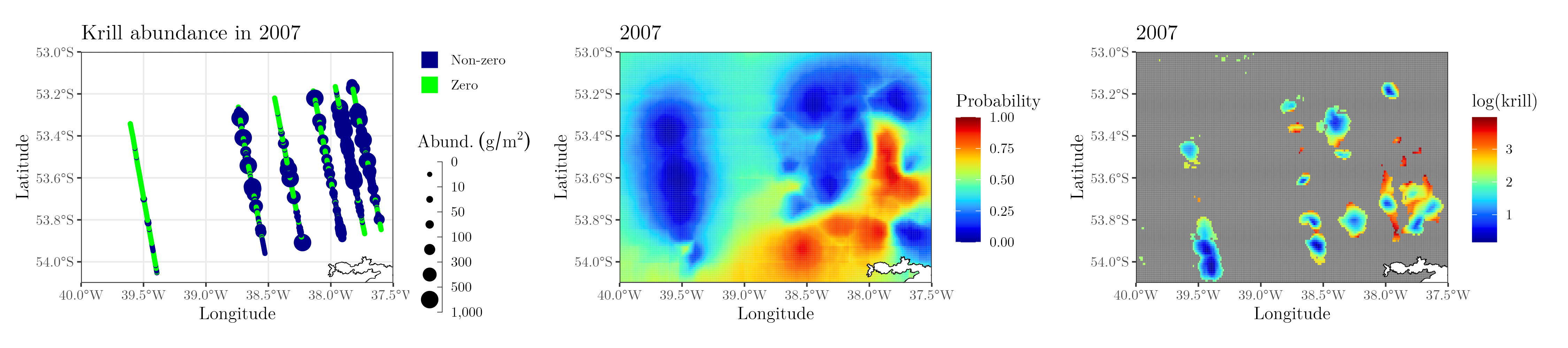}
    \caption{Year 2007. Left: observed acoustic krill biomass data. Middle: estimated probability of non-zeros. Right: predicted krill biomass ($\text{g}/\text{m}^2$), with predictions having a standard deviation greater than 3 (on the log scale) being masked out. The two rightmost plots are based on the mean of the corresponding predictive distributions.}
	\label{fig:all_pred_original_data_2007}
\end{figure}

\begin{figure}[!ht]
	\centering
	\includegraphics[width = 0.9\textwidth]{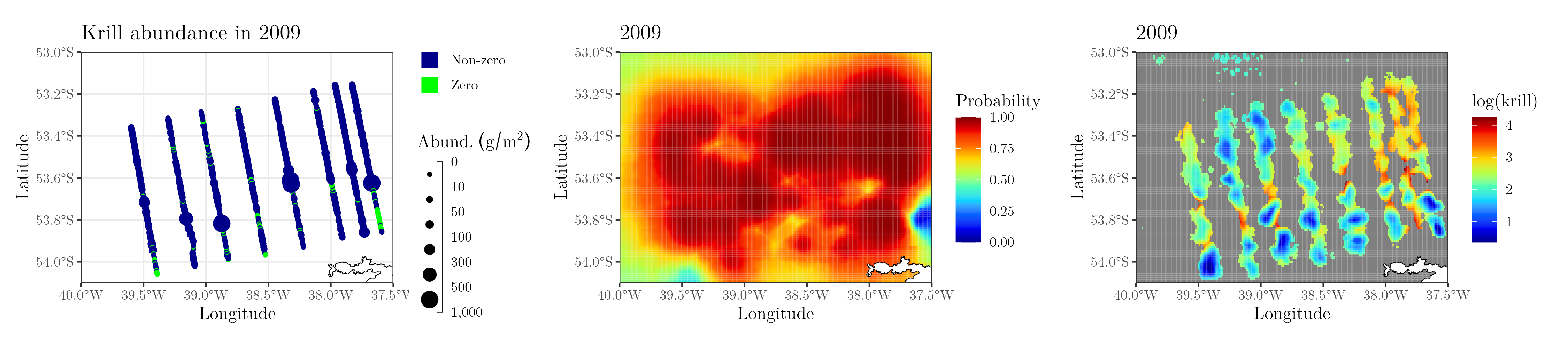}
    \caption{Year 2009. Left: observed acoustic krill biomass data. Middle: estimated probability of non-zeros. Right: predicted krill biomass ($\text{g}/\text{m}^2$), with predictions having a standard deviation greater than 3 (on the log scale) being masked out. The two rightmost plots are based on the mean of the corresponding predictive distributions.}
	\label{fig:all_pred_original_data_2009}
\end{figure}

\begin{figure}[!ht]
	\centering
	\includegraphics[width = 0.9\textwidth]{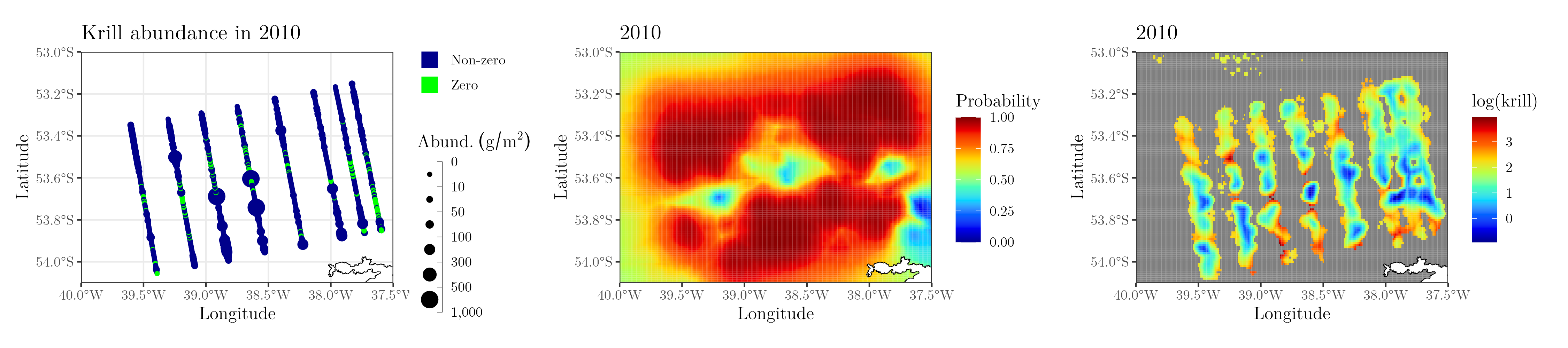}
    \caption{Year 2010. Left: observed acoustic krill biomass data. Middle: estimated probability of non-zeros. Right: predicted krill biomass ($\text{g}/\text{m}^2$), with predictions having a standard deviation greater than 3 (on the log scale) being masked out. The two rightmost plots are based on the mean of the corresponding predictive distributions.}
	\label{fig:all_pred_original_data_2010}
\end{figure}

\begin{figure}[!ht]
	\centering
	\includegraphics[width = 0.9\textwidth]{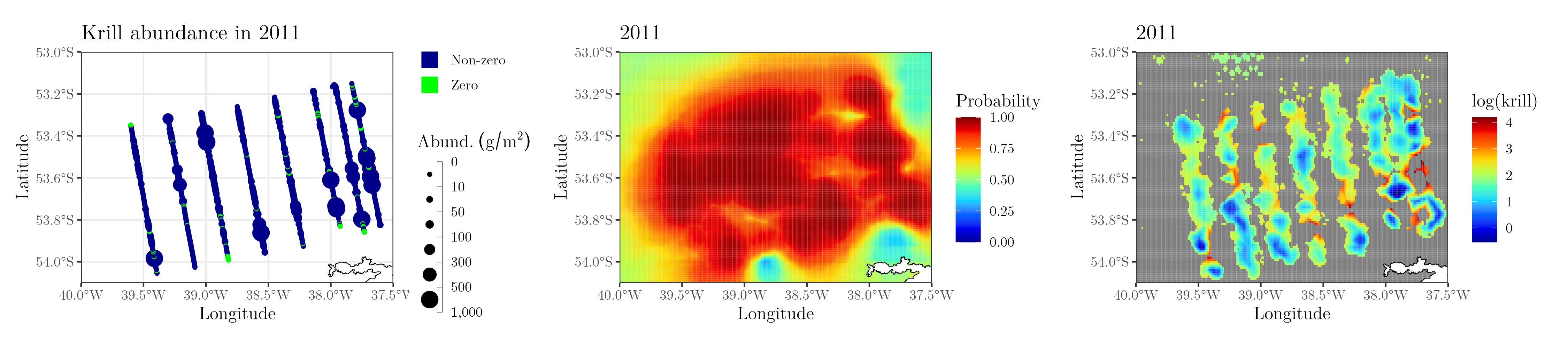}
    \caption{Year 2011. Left: observed acoustic krill biomass data. Middle: estimated probability of non-zeros. Right: predicted krill biomass ($\text{g}/\text{m}^2$), with predictions having a standard deviation greater than 3 (on the log scale) being masked out. The two rightmost plots are based on the mean of the corresponding predictive distributions.}
	\label{fig:all_pred_original_data_2011}
\end{figure}

\begin{figure}[!ht]
	\centering
	\includegraphics[width = 0.9\textwidth]{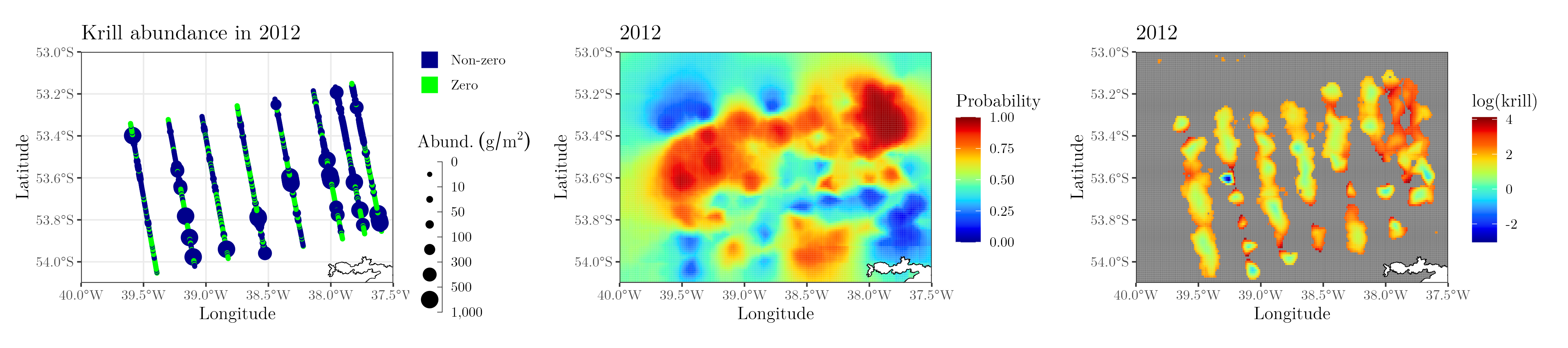}
    \caption{Year 2012. Left: observed acoustic krill biomass data. Middle: estimated probability of non-zeros. Right: predicted krill biomass ($\text{g}/\text{m}^2$), with predictions having a standard deviation greater than 3 (on the log scale) being masked out. The two rightmost plots are based on the mean of the corresponding predictive distributions.}
	\label{fig:all_pred_original_data_2012}
\end{figure}

\begin{figure}[!ht]
	\centering
	\includegraphics[width = 0.9\textwidth]{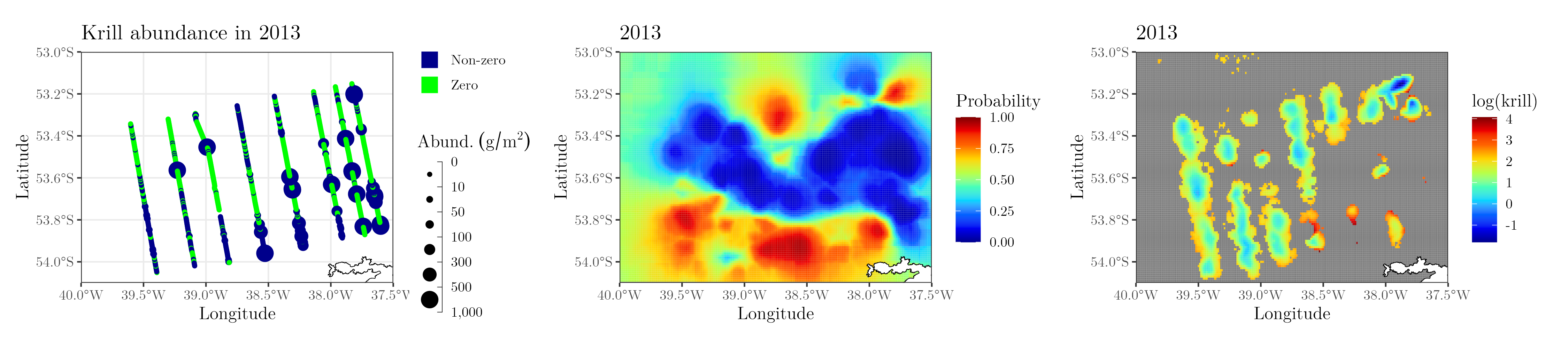}
    \caption{Year 2013. Left: observed acoustic krill biomass data. Middle: estimated probability of non-zeros. Right: predicted krill biomass ($\text{g}/\text{m}^2$), with predictions having a standard deviation greater than 3 (on the log scale) being masked out. The two rightmost plots are based on the mean of the corresponding predictive distributions.}
	\label{fig:all_pred_original_data_2013}
\end{figure}

\begin{figure}[!ht]
	\centering
	\includegraphics[width = 0.9\textwidth]{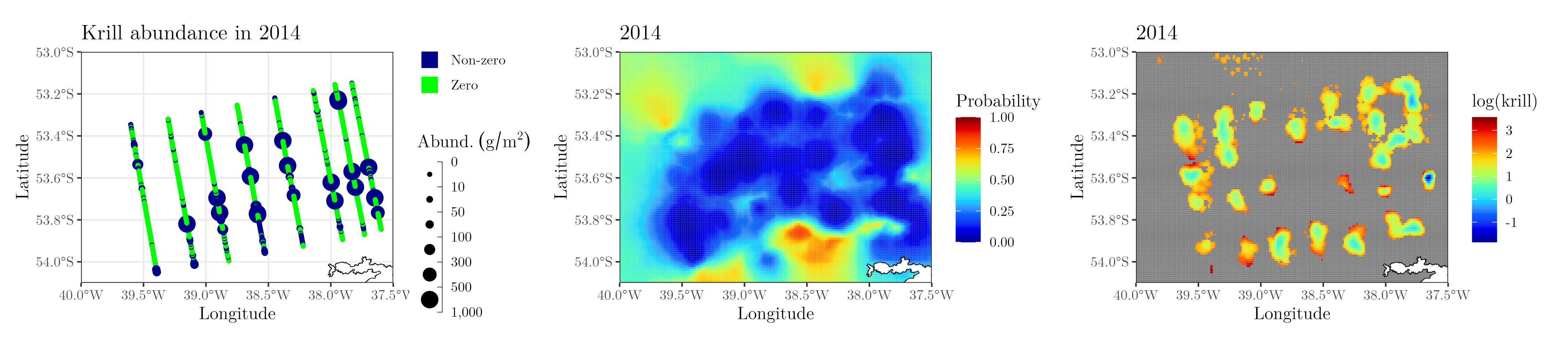}
    \caption{Year 2014. Left: observed acoustic krill biomass data. Middle: estimated probability of non-zeros. Right: predicted krill biomass ($\text{g}/\text{m}^2$), with predictions having a standard deviation greater than 3 (on the log scale) being masked out. The two rightmost plots are based on the mean of the corresponding predictive distributions.}
	\label{fig:all_pred_original_data_2014}
\end{figure}

\begin{figure}[!ht]
	\centering
	\includegraphics[width = 0.9\textwidth]{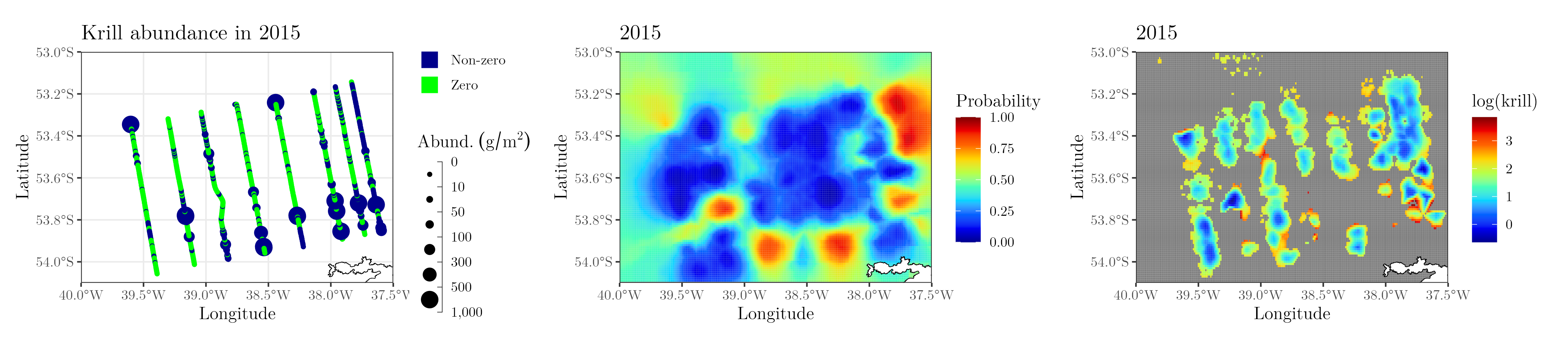}
    \caption{Year 2015. Left: observed acoustic krill biomass data. Middle: estimated probability of non-zeros. Right: predicted krill biomass ($\text{g}/\text{m}^2$), with predictions having a standard deviation greater than 3 (on the log scale) being masked out. The two rightmost plots are based on the mean of the corresponding predictive distributions.}
	\label{fig:all_pred_original_data_2015}
\end{figure}

\begin{figure}[!ht]
	\centering
	\includegraphics[width = 0.9\textwidth]{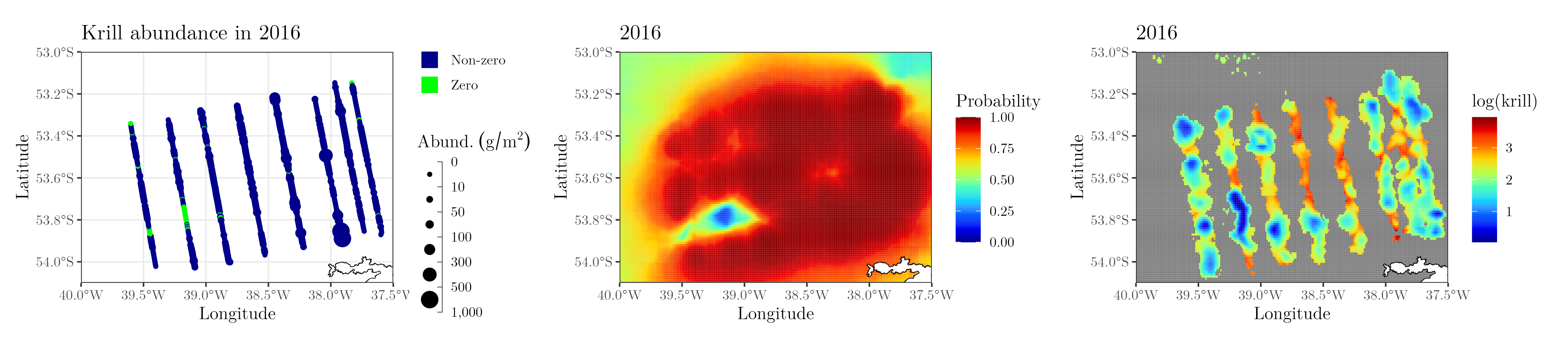}
    \caption{Year 2016. Left: observed acoustic krill biomass data. Middle: estimated probability of non-zeros. Right: predicted krill biomass ($\text{g}/\text{m}^2$), with predictions having a standard deviation greater than 3 (on the log scale) being masked out. The two rightmost plots are based on the mean of the corresponding predictive distributions.}
	\label{fig:all_pred_original_data_2016}
\end{figure}

\begin{figure}[!ht]
	\centering
	\includegraphics[width = 0.9\textwidth]{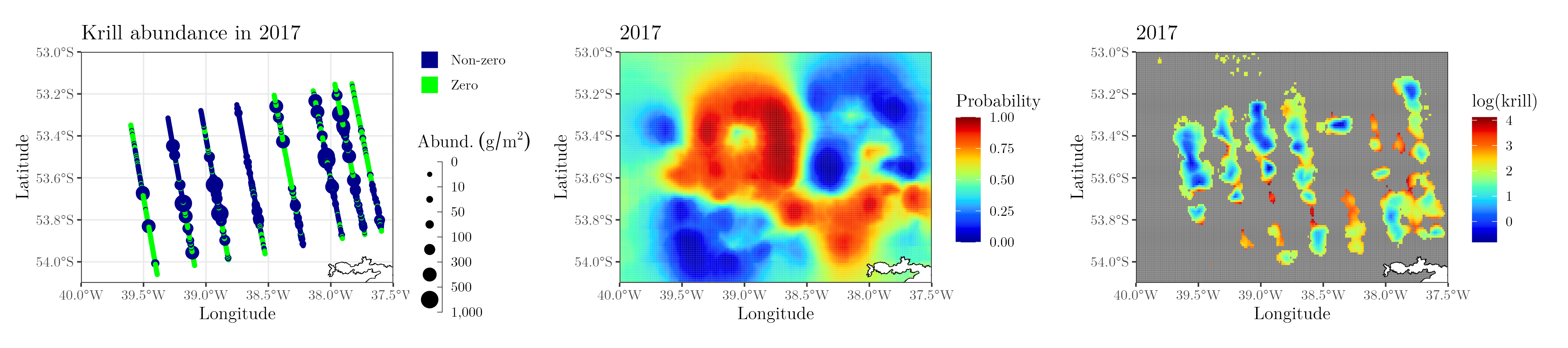}
    \caption{Year 2017. Left: observed acoustic krill biomass data. Middle: estimated probability of non-zeros. Right: predicted krill biomass ($\text{g}/\text{m}^2$), with predictions having a standard deviation greater than 3 (on the log scale) being masked out. The two rightmost plots are based on the mean of the corresponding predictive distributions.}
	\label{fig:all_pred_original_data_2017}
\end{figure}

\begin{figure}[!ht]
	\centering
	\includegraphics[width = 0.9\textwidth]{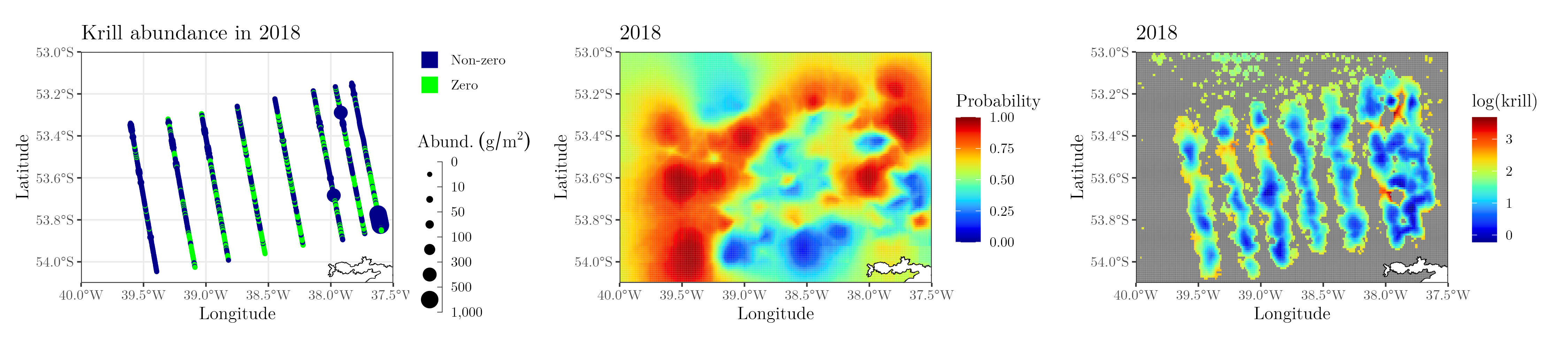}
    \caption{Year 2018. Left: observed acoustic krill biomass data. Middle: estimated probability of non-zeros. Right: predicted krill biomass ($\text{g}/\text{m}^2$), with predictions having a standard deviation greater than 3 (on the log scale) being masked out. The two rightmost plots are based on the mean of the corresponding predictive distributions.}
	\label{fig:all_pred_original_data_2018}
\end{figure}

\begin{figure}[!ht]
	\centering
	\includegraphics[width = 0.9\textwidth]{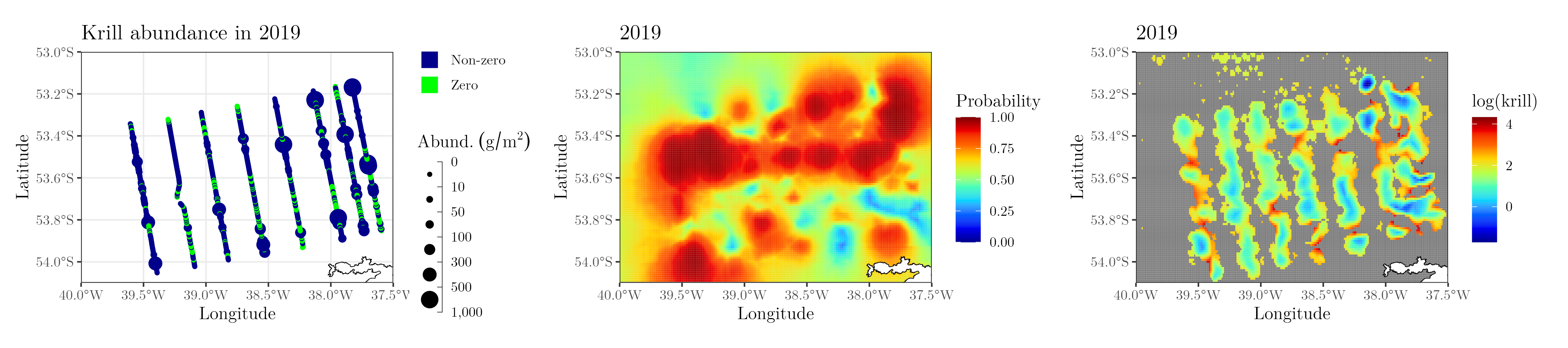}
    \caption{Year 2019. Left: observed acoustic krill biomass data. Middle: estimated probability of non-zeros. Right: predicted krill biomass ($\text{g}/\text{m}^2$), with predictions having a standard deviation greater than 3 (on the log scale) being masked out. The two rightmost plots are based on the mean of the corresponding predictive distributions.}
	\label{fig:all_pred_original_data_2019}
\end{figure}

Lastly, Figures \ref{fig:all_pred_original_data_1998}--\ref{fig:all_pred_original_data_2019} display the acoustic surveys and predicted values for both the presence-absence and positive biomass components for the years not shown in \refmain{Section 4.1}.

\newpage

\subsubsection{Sensitivity analysis} \label{sssec:S-sensitivy}

To evaluate the robustness of our hyperparameter estimates under a misspecified model, we simulate data from a Matérn model with normal inverse Gaussian (NIG) noise and attempt to retrieve key quantities, particularly $\kappa = \sqrt{8\nu}/\rho$, where $\rho$ denotes the range.

The NIG distribution is a continuous probability distribution belonging to the generalized hyperbolic family \citep{barndorff:1978}. It is particularly useful due to its flexibility in modelling asymmetry and heavy tails. For a random variable $X$ following an NIG distribution with parameters $\nu, \phantom{|}\sigma > 0$ and $\delta, \phantom{|}\mu \in \mathbb{R}$, the probability density function is given by
\begin{align} \label{eq:NIG}
    f(x; \delta, \mu, \sigma, \nu) = \frac{e^{\nu + \mu(x - \delta)/\sigma^2}}{\pi \sqrt{\nu \sigma^2 + (x - \delta)^2}} \sqrt{\nu \mu^2/\sigma^2 + \nu^2} K_1 \left( \sqrt{\left( \nu \sigma^2 + (x - \delta)^2 \right)\left( \mu^2/\sigma^4 + \nu/\sigma^2\right)} \right),
\end{align}
where $K_p(\cdot)$ is modified Bessel function of the second kind of order $p$. This parametrisation is the same as the one used in the \texttt{ngme2} package \citep{ngme2:2024}, where $\delta$, $\mu$, $\sigma$, and $\nu$ represent the parameters for location, skewness, scaling, and shape, respectively. 

In this section, we simulate data from a Matérn model with NIG noise in the unit square, i.e., $\mathcal{X} = [0, 1] \times [0, 1]$, with $m$ replicates. Specifically, we consider the following model. Let $y(s) = (y(s_1), \cdots , y(s_n))$ represent the data set observed at locations $s = (s_1, \cdots, s_n) \subset \mathcal{X}$, then
\begin{align} \label{eq:Gaussian-NIG}
    y(s) &= \beta_0 + \phi(s) +\epsilon(s), \text{ s.t. } \epsilon(s) \overset{\text{i.i.d.}}{\sim} \text{Normal}(0, \sigma^2_\epsilon) \\
    \phi(s)|\Lambda(s) &= \delta + \mu \Lambda(s) + \sigma \sqrt{\Lambda(s)} \xi(s) \nonumber \\ 
    %\text{Gaussian Process}(\delta + \mu \Lambda(s), \Sigma_{\phi}) \nonumber \\
    \Lambda(s) &\overset{\text{i.i.d.}}{\sim} \text{Inverse Gaussian}(\nu, \nu), \nonumber
\end{align}
where $\xi(s)$ is Gaussian Process with mean $0$ and a Matérn kernel given by $r(h; \theta)$, such that $\theta = (\nu_{\phi}, \kappa_{\phi})$.

\begin{figure}[!ht]
	\centering
	\includegraphics[width = 1\textwidth]{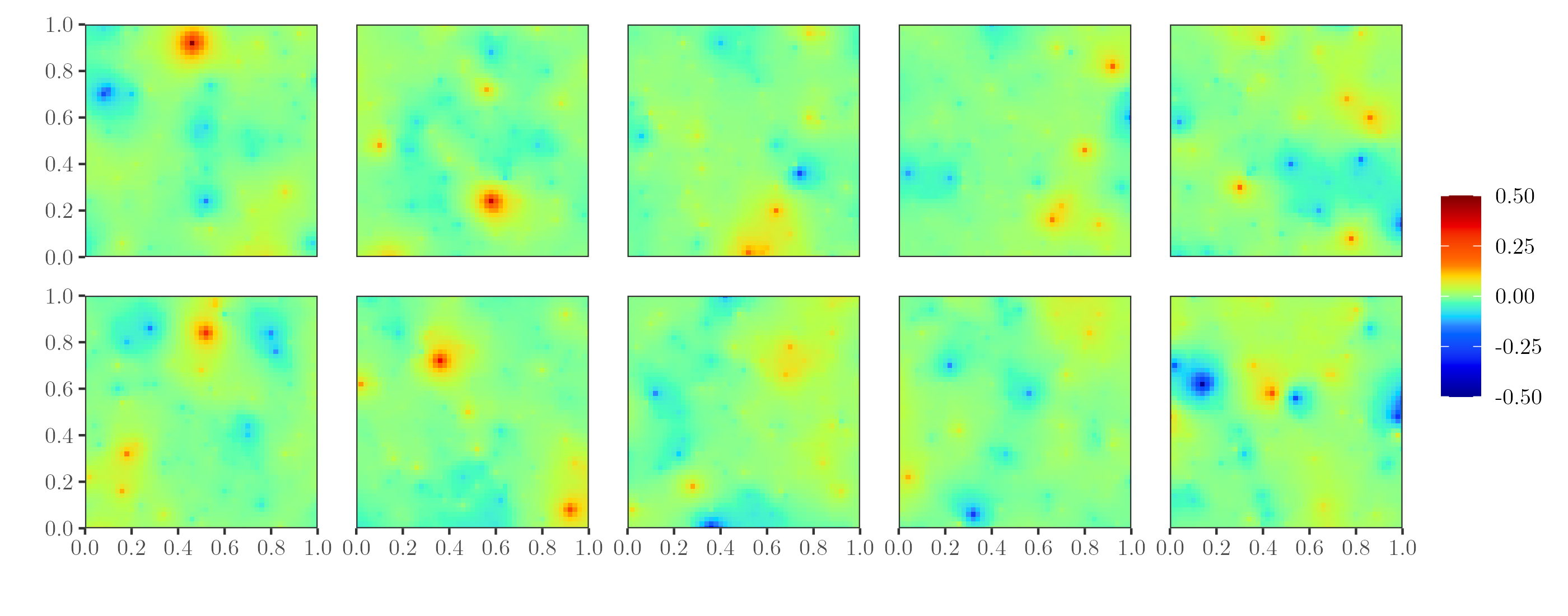}
    \caption{10 simulations from a process defined in $\phantom{|}[0, 1] \times [0, 1]\phantom{|}$ by a Matérn kernel and NIG noise.}
	\label{fig:nig_simulation}
\end{figure}

In particular, we set $\beta_0 = 0$, $\nu_{\phi}= 1$, $\kappa_{\phi} = 10$, $\sigma = 1$, $\delta = 0$, $\mu = 0$, $\nu = 10$, and $\sigma_{\epsilon} = 0.01$ (i.e., for simplicity, we assume the observational noise to be very small), with $m = 10$ replicates. Figure \ref{fig:nig_simulation} shows all the simulated surfaces $\phi(s)$.

Based on the realizations shown in Figure \ref{fig:nig_simulation}, we randomly sample the observed area at $n = 50$, $100$, $500$, and $1{,}000$ locations and then estimate the Matérn hyperparameters using the correctly specified model (i.e., NIG noise) and a misspecified model with Gaussian noise. Table \ref{tab:S-est_coeff_nig} presents the obtained estimates for all combinations of fitted models and sample sizes. Inference was made using the \texttt{ngme2} package \citep{ngme2:2024}. 

\begin{table}[!ht]
	\caption{Estimated hyperparameters for the Matérn structure based on data generated as described in Equation \eqref{eq:Gaussian-NIG} for a correctly specified model (NIG) and a misspecified model (Gaussian). The parameter estimates are based on the mean and standard deviation of the final 2,000 samples, obtained after 10,000 iterations with a burn-in of 8,000. Note that $\nu_{\phi}$ is not estimated but fixed.}
    \resizebox{\textwidth}{!}{%
    \centering
    \begin{tabular}{ c | c | c | c | c | c | c | c | c | c }
    \multirow{2}{*}{Model} & \multirow{2}{*}{Sample size} & \multicolumn{3}{c |}{Parameters} & \multirow{2}{*}{Model} & \multirow{2}{*}{Sample size} & \multicolumn{3}{c}{Parameters} \\ \cline{3-5} \cline{8-10}
    & & & True & Estimated & & & & True & Estimated \\ \hline
    \multirow{8}{*}{NIG}      & \multirow{2}{*}{$\phantom{1{,}1}50$} & $\sigma$        & $\phantom{0}$1 & $\phantom{1}2.15 ~ (0.96)$ & \multirow{8}{*}{Gaussian} & \multirow{2}{*}{$\phantom{1{,}1}50$} & $\sigma$        & $\phantom{0}$1 & $\phantom{1}1.28 ~ (0.20)$ \\
                              &                                      & $\kappa_{\phi}$ & $10$           & $10.58 ~ (2.70)$           &                           &                                      & $\kappa_{\phi}$ & $10$           & $12.36 ~ (1.93)$           \\ \cline{2-5} \cline{7-10}
                              & \multirow{2}{*}{$\phantom{1{,}}100$} & $\sigma$        & $\phantom{0}$1 & $\phantom{1}1.03 ~ (0.34)$ &                           & \multirow{2}{*}{$\phantom{1{,}}100$} & $\sigma$        & $\phantom{0}$1 & $\phantom{1}1.04 ~ (0.17)$ \\ 
                              &                                      & $\kappa_{\phi}$ & $10$           & $\phantom{1}9.56 ~ (1.77)$ &                           &                                      & $\kappa_{\phi}$ & $10$           & $11.02 ~ (1.78)$           \\ \cline{2-5} \cline{7-10}
                              & \multirow{2}{*}{$\phantom{1{,}}500$} & $\sigma$        & $\phantom{0}$1 & $\phantom{1}1.01 ~ (0.17)$ &                           & \multirow{2}{*}{$\phantom{1{,}}500$} & $\sigma$        & $\phantom{0}$1 & $\phantom{1}1.07 ~ (0.08)$ \\
                              &                                      & $\kappa_{\phi}$ & $10$           & $10.21 ~ (0.78)$           &                           &                                      & $\kappa_{\phi}$ & $10$           & $11.08 ~ (1.03)$           \\ \cline{2-5} \cline{7-10}
                              & \multirow{2}{*}{$1{,}000$}           & $\sigma$        & $\phantom{0}$1 & $\phantom{1}1.09 ~ (0.20)$ &                           & \multirow{2}{*}{$1{,}000$}           & $\sigma$        & $\phantom{0}$1 & $\phantom{1}1.00 ~ (0.05)$ \\
                              &                                      & $\kappa_{\phi}$ & $10$           & $\phantom{1}9.71 ~ (0.66)$ &                           &                                      & $\kappa_{\phi}$ & $10$           & $10.30 ~ (0.78)$           \\ 
    \end{tabular}
    }
	\label{tab:S-est_coeff_nig}
\end{table}

From Table \ref{tab:S-est_coeff_nig}, we observe that in the NIG model, $\sigma$ is more challenging to estimate, particularly for smaller sample sizes (e.g., $n = 50$), which is a known issue due to the parametrisation in Equation \eqref{eq:NIG}. On the other hand, $\kappa_{\phi}$ appears to be reasonably well-estimated. In contrast, when analysing the Gaussian model, although the scaling parameter is well-recovered, $\kappa_{\sigma}$ is slightly overestimated (by 5–10\%, when $n \geq 100$). Therefore, in a misspecified setting where the observed process exhibits peaks and valleys, the range parameter $\rho_{\phi} = \sqrt{8\nu_{\phi}}/\kappa_{\phi}$ may be underestimated---although the bias is expected to remain modest if the distribution tails of the underlying data are not excessively heavy.

\subsection{Aggregated Spatial Modelling} \label{ssec:S-agg_sup}

In \refmain{Section 4.2}, we implemented a spatial Hurdle-Gamma model (as stated in \refmain{Section 3.1.1}) with linear predictors specified as follows
\begin{align} \label{eq:lp_z_agg}
    \text{logit}(\pi_{i}) = ~ 
    \beta_0^z + 
    \beta_1^z \texttt{chlor}_i +
    \beta_2^z \texttt{pot\_temp}_i + 
    \beta_3^z \texttt{speed\_sat}_i +
    \beta_4^z \texttt{surf\_temp}_i +
    \beta_5^z \texttt{res\_time}_i +
    \beta_6^z \texttt{mass\_flux}_i + 
    \psi_{i},
\end{align}
and
\begin{align} \label{eq:lp_y_agg}
    \log(\mu_{i}) = ~ 
    \beta_0^y + 
    \beta_1^y \texttt{depth}_i + 
    \beta_2^y \texttt{salinity}_i + 
    \beta_3^y \texttt{surf\_temp}_i + 
    \gamma \cdot \psi_{i} + \xi_{i},
\end{align}
where $\gamma$ is a ``copy'' factor, and $\psi_{i}$ (similarly, $\xi_{i}$) is a spatial random effect modelled as a Gaussian process with a Matérn covariance structure, characterized by the range parameter $\rho_{\psi}$ and marginal variance $\sigma_{\psi}^2$.

\begin{figure}[!ht]
	\centering
	\includegraphics[width = 0.95\textwidth]{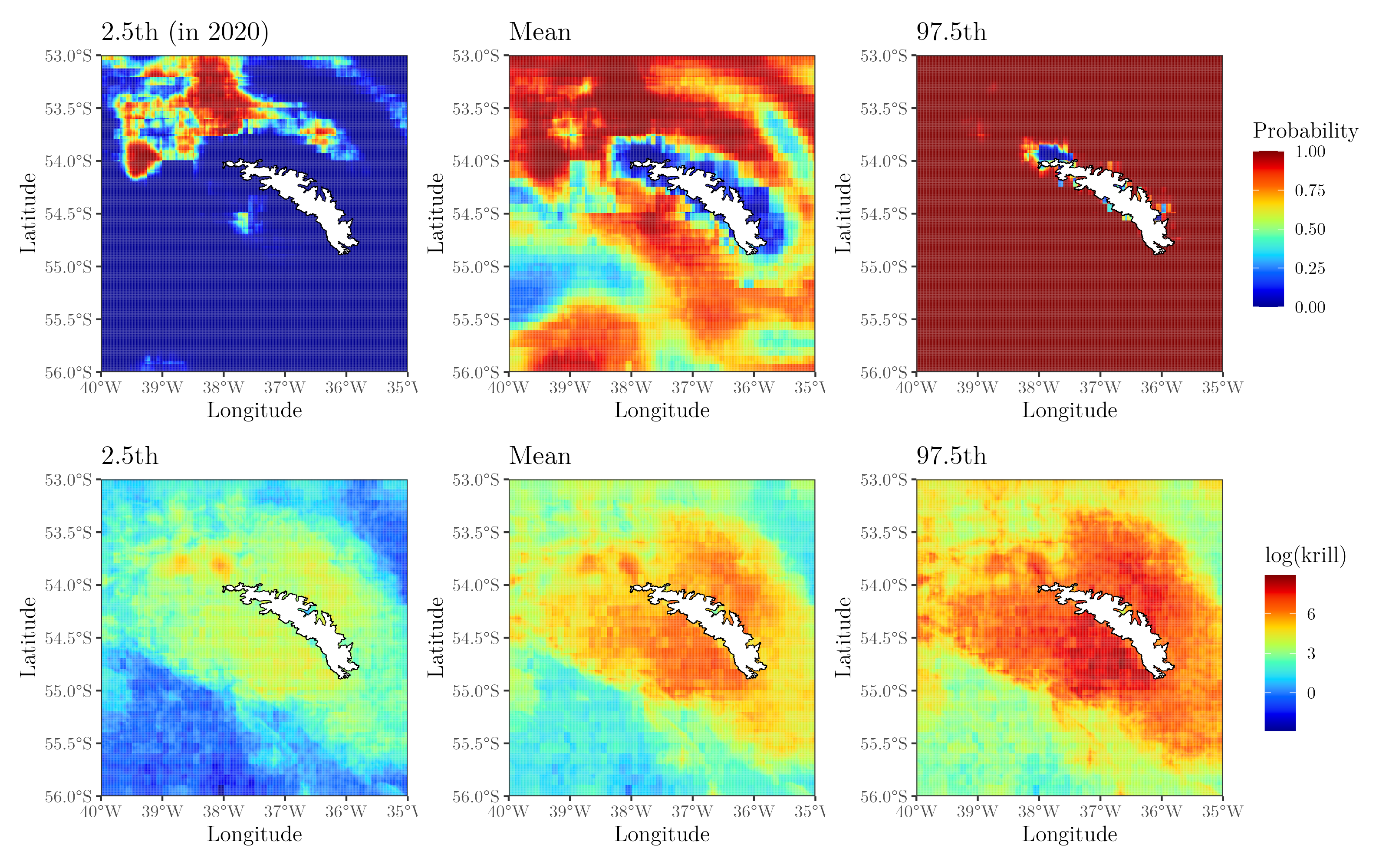}
    \caption{Year 2020. Top row: predicted probability of non-zero krill biomass (posterior mean) along with associated uncertainty. Bottom row: predicted krill biomass (posterior mean) along with associated uncertainty. Krill biomass is in $\text{g}/\text{m}^2$.}
	\label{fig:S-all_pred_unc}
\end{figure}

As before, for numerical stability, all covariates were re-scaled. The estimated coefficients are shows in Table \ref{tab:S-est_coeff_agg}. Figure \ref{fig:S-all_pred_unc} shows the predicted krill biomass for 2020 along with the associated prediction uncertainty, represented by the $2.5{\text{th}}$ and $97.5{\text{th}}$ quantiles. The same figure also shows the probabilities of observing non-zero krill biomass.

\begin{table}[!ht]
	\caption{Estimated parameters (with standard deviation and a 95\% equal-tail credible interval) for the spatial model fitted for the aggregated data.}
    \resizebox{\textwidth}{!}{%
    \centering
    \begin{tabular}{ c | c | c | c | c | c | c | c | c | c | c | c }
    Parameter & Mean & SD & 95\% equal-tail CI & Parameter & Mean & SD & 95\% equal-tail CI & Parameter & Mean & SD & 95\% equal-tail CI \\ \hline
    $\beta_{0}^z$  & $\phantom{-}6.035$ & $1.673$ & $(\phantom{-}2.755; \phantom{-}9.315)$ & $\beta_{0}^y$  & $\phantom{-}3.265$ & $0.249$ & $(\phantom{-}2.777; 3.754)$ & $\rho_{\psi}$   & $50.043$           & $28.115$           & $(13.313; 120.340)$ \\
    $\beta_{1}^z$  & $-1.323$           & $0.800$ & $(-2.891; \phantom{-}0.246)$           & $\beta_{1}^y$  & $\phantom{-}1.502$ & $0.336$ & $(\phantom{-}0.845; 2.160)$ & $\sigma_{\psi}$ & $\phantom{0}3.788$ & $\phantom{0}0.943$ & $(\phantom{0}2.148; \phantom{00}5.821)$ \\
    $\beta_{2}^z$  & $-1.887$           & $0.532$ & $(-2.929; -0.846)$                     & $\beta_{2}^y$  & $\phantom{-}0.681$ & $0.350$ & $(-0.005; 1.367)$           & $\rho_{\xi}$    & $12.786$           & $\phantom{0}5.832$ & $(\phantom{0}5.398; \phantom{0}27.757)$ \\
    $\beta_{3}^z$  & $-2.114$           & $0.838$ & $(-3.756; -0.472)$                     & $\beta_{3}^y$  & $\phantom{-}0.590$ & $0.094$ & $(\phantom{-}0.405; 0.774)$ & $\sigma_{\xi}$  & $\phantom{0}1.045$ & $\phantom{0}0.815$ & $(\phantom{0}0.163; \phantom{00}3.177)$ \\
    $\beta_{4}^z$  & $\phantom{-}1.138$ & $0.458$ & $(\phantom{-}0.240, \phantom{-}2.036)$ &                &                    &         &                             & $\gamma$        & $-0.144$           & $\phantom{0}0.059$ & $(-0.264; -0.031)$ \\
    $\beta_{5}^z$  & $\phantom{-}0.430$ & $1.071$ & $(-1.668, \phantom{-}2.529)$           &                &                    &         &                             &&&& \\
    $\beta_{6}^z$  & $-0.653$           & $1.114$ & $(-2.836, \phantom{-}1.530)$           &                &                    &         &                             &&&& \\
    \end{tabular}
    }
	\label{tab:S-est_coeff_agg}
\end{table}

Table \ref{tab:S-est_coeff_agg} and the reduced linear predictors in Equations \eqref{eq:lp_z_agg} and \eqref{eq:lp_y_agg} show that the spatial-only analysis resulted in a much simpler specification than the disaggregated spatio-temporal model discussed in Section \ref{ssec:S-disagg_sup}. This parsimony is a direct consequence of the step-wise forward-selection routine, which retained only the covariates that improved model fit.

\subsubsection{Alternative Random Effects} \label{sssec:S-other-re}

Alternatively, we considered different random effect structures and, for each class of models, performed stepwise forward selection of covariates using the Watanabe-Akaike Information Criterion (WAIC) \citep{watanabe:2013, gelman:2014}. In particular, we considered three additional models, which are described as follows 
\begin{enumerate} [label=\arabic*., left=0pt, align=left]
    \item No random effects
    \vspace{-0.75\baselineskip}
    \begin{align} 
    \text{logit}(\pi_{i}) &= \beta_0^z + \beta_1^z\texttt{chlor}_{i} \label{eq:alternative-model-1_1} \\
    \log(\mu_{i}) &= \beta_0^y + \beta_1^y\texttt{depth}_{i} + \beta_2^y\texttt{speed}_{i} + \beta_3^y\texttt{surf\_temp}_{i} + \beta_3^y\texttt{mass\_flux}_{i} + \beta_4^y\texttt{density\_drif}_{i} \label{eq:alternative-model-1_2} 
    \end{align}
    \vspace{-1.5\baselineskip}
    \item Independent random effects
    \vspace{-0.75\baselineskip}
    \begin{align} 
    \text{logit}(\pi_{i}) &= \beta_0^z + \beta_1^z\texttt{salinity}_{i} + \psi_{i} \label{eq:alternative-model-2_1} \\
    \log(\mu_{i}) &= \beta_0^y + \beta_1^y\texttt{depth}_{i} + \beta_2^y\texttt{surf\_temp}_{i} + \beta_3^y\texttt{expect\_freq}_{i} + \xi_{i} \label{eq:alternative-model-2_2} 
    \end{align}
    \vspace{-1.5\baselineskip}
    \item Shared random effects
    \vspace{-0.75\baselineskip}
    \begin{align} 
    \text{logit}(\pi_{i}) &= \beta_0^z + \beta_1^z\texttt{expect\_freq}_{i} + \beta_2^z\texttt{res\_time}_{i} + \psi_{i} \label{eq:alternative-model-3_1} \\
    \log(\mu_{i}) &= \beta_0^y + \beta_1^y \texttt{depth}_{i} + \beta_2^y \texttt{chlor}_{i} + \beta_3^y \texttt{pot\_temp}_{i} + \beta_4^y \texttt{surf\_temp}_{i} + \beta_5^y \texttt{speed\_drif}_{i} + \gamma \cdot \psi_{i} \label{eq:alternative-model-3_2}
    \end{align}
\end{enumerate}
However, as shown in Table \ref{tab:S-comparing-alt-models}, the model specified by Equations \eqref{eq:lp_z_agg} and \eqref{eq:lp_y_agg} was selected based on the Deviance Information Criterion (DIC) \citep{spiegelhalter:2002} and the WAIC.

\begin{table}[!ht]
	\caption{Computed DIC and WAIC for alternative spatial models fitted based on the aggregated acoustic krill biomass data. All criterion are negatively oriented, meaning that smaller values are better.}
    % \resizebox{\textwidth}{!}{%
    \centering
    \begin{tabular}{ l | c | c } 
    Model & DIC & WAIC \\ \hline
    Equations \eqref{eq:lp_z_agg} and \eqref{eq:lp_y_agg}                           & $2911.852$ & $2937.433$ \\
    Equations \eqref{eq:alternative-model-1_1} and \eqref{eq:alternative-model-1_2} & $2927.787$ & $2956.351$ \\
    Equations \eqref{eq:alternative-model-2_1} and \eqref{eq:alternative-model-2_2} & $2921.207$ & $2948.681$ \\
    Equations \eqref{eq:alternative-model-3_1} and \eqref{eq:alternative-model-3_2} & $3046.863$ & $3058.399$ \\
    \end{tabular}
    % }
	\label{tab:S-comparing-alt-models}
\end{table}

\subsubsection{Net haul data} \label{sssec:S-krillbase_time}

\begin{figure}[!ht]
	\centering
	\includegraphics[width = 0.90\textwidth]{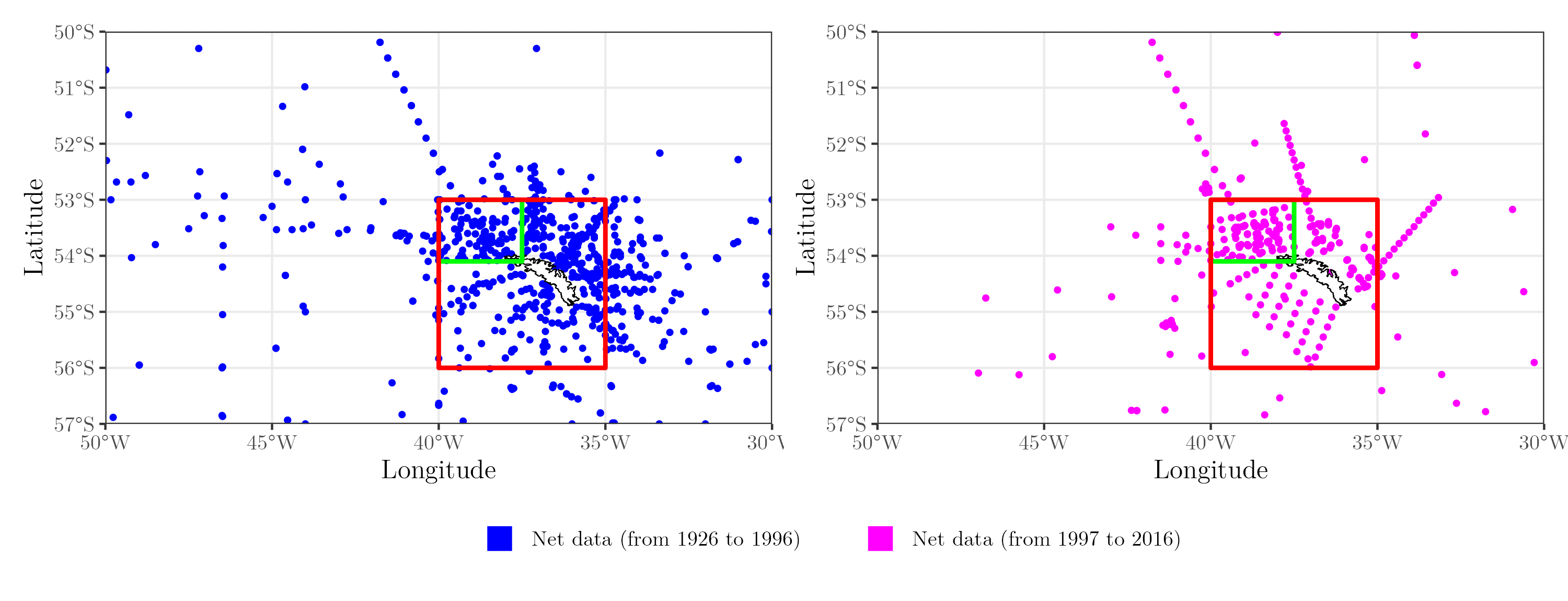}
    \caption{Sampling locations of net haul data from KRILLBASE collected between 1926 and 1996 (left panel) and 1997 and 2016 (right panel). The green box indicates the region where acoustic data were collected, while the red box marks the area with sufficient net haul data to serve as a validation set for model predictions.}
	\label{fig:net-haul-times}
\end{figure}

Figure \ref{fig:net-haul-times} shows the locations of net haul data (KRILLBASE) observations, split into two periods: 1926--1996 and 1997--2016, where the latter period matches the time window of the acoustic krill data.

% \bibliographystyle{cell}
% \bibliography{references}